\documentclass[11pt]{article}
\usepackage[utf8]{inputenc}
\usepackage{amsfonts}
\usepackage{amsmath}
\usepackage{amssymb}
\usepackage{indentfirst}
\usepackage{graphicx}
\usepackage{cite}
\usepackage{colortbl}
\usepackage{hyperref}
\usepackage{color}
\usepackage{xcolor}

\usepackage{microtype}
\usepackage[normalem]{ulem}

\numberwithin{equation}{section}
\allowdisplaybreaks
\makeatletter

\begin{document}

\begin{titlepage}
\vspace{3cm}

\baselineskip=24pt

\begin{center}
\textbf{\LARGE{Three-dimensional exotic Newtonian supergravity theory with cosmological constant}}
\par\end{center}{\LARGE \par}

\begin{center}
	\vspace{1cm}
	\textbf{Patrick Concha}$^{\ast}$,
    \textbf{Lucrezia Ravera}$^{\star, \ddag}$,
	\textbf{Evelyn Rodríguez}$^{\dag}$
	\small
	\\[5mm]
    $^{\ast}$\textit{Departamento de Matemática y Física Aplicadas, }\\
	\textit{ Universidad Católica de la Santísima Concepción, }\\
\textit{ Alonso de Ribera 2850, Concepción, Chile.}
	\\[2mm]
	$^{\star}$\textit{DISAT, Politecnico di Torino, }\\
	\textit{ Corso Duca degli Abruzzi 24, 10129 Torino, Italy.}
	\\[3mm]
	$^{\ddag}$\textit{INFN, Sezione di Torino, }\\
	\textit{ Via P. Giuria 1, 10125 Torino, Italy.}
	\\[3mm]
	$^{\dag}$\textit{Departamento de Física, Universidad del Bío-Bío, }\\
	\textit{Avenida Collao 1202, Casilla 5-C, Concepción, Chile}
	 \\[5mm]
	\footnotesize
	\texttt{patrick.concha@ucsc.cl},
    \texttt{lucrezia.ravera@polito.it},
	\texttt{everodriguezd@gmail.com},
	\par\end{center}
\vskip 26pt
\begin{abstract}
We present a supersymmetric extension of the exotic Newtonian Chern-Simons gravity theory in three spacetime dimensions. The underlying new non-relativistic superalgebra is obtained by expanding the $\mathcal{N}=2$ AdS superalgebra and can be written as two copies of the enhanced Nappi-Witten algebra, one of which is augmented by supersymmetry. We show that the exotic Newtonian superalgebra allows to introduce a cosmological constant to the extended Newtonian supergravity. Interestingly, the obtained supergravity action contains the extended Newton-Hooke supergravity as a sub-case.
\end{abstract}
\end{titlepage}\newpage {} {\baselineskip=12pt \tableofcontents{}}

\section{Introduction}

There has been a growing interest in exploring Newtonian (super)gravity theories due to their use in strongly coupled condensed matter systems \cite{Son:2008ye,Balasubramanian:2008dm,Kachru:2008yh,Bagchi:2009my,Bagchi:2009pe,Christensen:2013lma,Christensen:2013rfa,Hartong:2014oma,Hartong:2014pma,Hartong:2015wxa,Taylor:2015glc} and non-relativistic effective field theories \cite{Hoyos:2011ez,Son:2013rqa,Abanov:2014ula,Geracie:2015dea,Gromov:2015fda}. The construction of Newtonian gravity, describing the physical gravitational force at non-relativistic level, requires to consider the so-called Newton-Cartan geometry \cite{Cartan1,Cartan2}. Such geometrical framework is necessary to covariantize the Poisson equation of Newtonian gravity. Nevertheless, a principle action for Newtonian gravity was recently presented in \cite{Hansen:2018ofj} which has required to extend the Bargmann algebra \cite{LL,Grigore:1993fz,Bose:1994sj,Duval:2000xr,Jackiw:2000tz,Papageorgiou:2009zc,Grassie:2020dga, Bergshoeff:2016lwr} by including three additional generators. Subsequently, a three-dimensional Chern-Simons (CS) action has been constructed in \cite{Ozdemir:2019orp} which is invariant under a central extension of the symmetry group that leaves the recently constructed Newtonian gravity action invariant. The novel symmetry has been denoted as extended Newtonian algebra and can be recovered by means of a contraction of a bi-metric model being the sum of Einstein gravity in the Lorentzian and Euclidean signatures. Interestingly, unlike the Newtonian gravity of \cite{Hansen:2018ofj}, the matter coupling of the extended Newtonian gravity theory admits backgrounds with non-trivial curvature whenever matter is present, similarly to the matter-coupled extended Bargmann gravity \cite{Bergshoeff:2016lwr}.

The introduction of a cosmological constant in non-relativistic gravity theories is done considering the Newton-Hooke symmetry \cite{Bacry:1968zf, Aldrovandi:1998im, Gibbons:2003rv, Brugues:2006yd, Alvarez:2007fw, Duval:2011mi, Duval:2016tzi}. However, an extension of the extended Newton-Hooke algebra \cite{Papageorgiou:2010ud, Hartong:2016yrf} is needed to include a cosmological constant to the extended Newtonian gravity theory \cite{Concha:2019dqs}. The novel symmetry is denoted as exotic Newtonian algebra and can be seen as an enhanced Bargmann-Newton-Hooke algebra \cite{Bergshoeff:2020fiz}. Both extended and exotic Newtonian gravity theories can be recovered as the non-relativistic limit of the coadjoint $\text{Poincaré}\,\oplus\,\mathfrak{u}\left(1\right)^2$ and coadjoint $\text{AdS}\,\oplus\,\mathfrak{u}\left(1\right)^2$ gravity theories \cite{Bergshoeff:2020fiz}.

Supersymmetric extensions of three-dimensional non-relativistic gravity models have been recently approached in \cite{Andringa:2013mma, Bergshoeff:2015ija, Bergshoeff:2016lwr} and subsequently studied in \cite{Ozdemir:2019orp, deAzcarraga:2019mdn, Ozdemir:2019tby, Concha:2019mxx, Concha:2020tqx, Concha:2020eam}. In particular, a CS action based on the supersymmetric extension of the extended Newtonian algebra has been presented in \cite{Ozdemir:2019orp}. Although a cosmological constant has been accommodated in a non-relativistic supergravity theory through the extended Newton-Hooke superalgebra \cite{Ozdemir:2019tby}, the possible supersymmetric extensions of the exotic Newtonian gravity remain unexplored. Unlike bosonic non-relativistic gravity, the construction of an action based on a non-relativistic superalgebra is non-trivial and requires the introduction of additional bosonic generators.  Furthermore, the non-relativistic limit is often ambiguous when supercharges are present. One way to circumvent this difficulty is through the expansion method based on Maurer-Cartan forms \cite{deAzcarraga:2002xi} and semigroups \cite{Izaurieta:2006zz}, which have proved to be useful to obtain known and new non-relativistic supergravity theories from relativistic ones \cite{deAzcarraga:2019mdn, Ozdemir:2019tby, Concha:2020tqx}.

In this work, we present a supersymmetric extension of the three-dimensional exotic Newtonian CS gravity introduced in \cite{Concha:2019dqs} by applying the semigroup expansion ($S$-expansion) \cite{Izaurieta:2006zz} to the $\mathcal{N}=2$ AdS supergravity theory. The motivation to consider the CS formalism in three spacetime dimensions is twofold. On one hand, three-dimensional CS gravity can be seen as a toy model to approach higher-dimensional theories. On the other hand, the construction of a non-relativistic (super)gravity action is much simpler and affordable through the CS formalism which allows us to write a gauge-invariant action. The novel Newtonian supergravity theory is invariant under an exotic Newtonian superalgebra which can be written as two copies of the enhanced Nappi-Witten algebra \cite{Bergshoeff:2020fiz, Concha:2020ebl}, one of which is augmented by supersymmetry.  Interestingly, the extended Newton-Hooke supergravity action appears as a sub-case of the exotic Newtonian supergravity theory. Moreover, we show that the extended Bargmann and extended Newtonian supergravities are recovered in the vanishing cosmological constant limit (i.e. the flat limit).

The paper is organized as follows: In Section \ref{sec2} we review the exotic Newtonian algebra and the corresponding three-dimensional CS action. Sections \ref{sec3} and \ref{sec4} contain our main results. In Section \ref{sec3}, we present an exotic Newtonian superalgebra by considering an $S$-expansion of the $\mathcal{N}=2$ super AdS algebra. In Section \ref{sec4}, we construct the exotic Newtonain CS supergravity action and analyze its vanishing cosmological constant limit. Section \ref{sec5} is devoted to discussion and future outlook.


\section{Exotic Newtonian gravity in three spacetime dimensions}\label{sec2}

In this section, we briefly review the three-dimensional CS exotic Newtonian gravity theory presented in \cite{Concha:2019dqs}. The exotic Newtonian algebra, also denoted as extended post-Newtonian algebra \cite{Gomis:2019nih}, is spanned by the set of generators $\{J,G_a,H,P_a,M,S,B_a,T_a\}$ along with two central charges $Y$ and $Z$. The non-vanishing commutation relations of the exotic Newtonian algebra are given by
\begin{eqnarray}
\left[ J,G_{a}\right] &=&\epsilon _{ab}G_{b}\,, \qquad %
\left[ G_{a},G_{b}\right] =-\epsilon _{ab}S\,, \qquad %
\left[ H,G_{a}\right] =\epsilon _{ab}P_{b}\,,  \notag
\\
\left[ J,P_{a}\right] &=&\epsilon _{ab}P_{b}\,,\qquad \, %
\left[ G_{a},P_{b}\right] =-\epsilon _{ab}M\,,\quad \ \ %
\left[ H,B_{a}\right] =\epsilon _{ab}T_{b}\,,  \notag
\\
\left[ J,B_{a}\right] &=&\epsilon _{ab}B_{b}\,,\qquad %
\left[ G_{a},B_{b}\right] =-\epsilon _{ab}Z\,,\qquad \,  %
\left[ S,G_{a}\right] =\epsilon _{ab}B_{b}\,,
\notag \\
\left[ J,T_{a}\right] &=&\epsilon _{ab}T_{b}\,,\qquad \  %
\left[ G_{a},T_{b}\right] =-\epsilon _{ab}Y\,,\qquad \    %
\left[ S,P_{a}\right] =\epsilon _{ab}T_{b}\,,
\notag \\
\left[ M,G_{a}\right] &=&\epsilon _{ab}T_{b}\,,\qquad \  %
\left[ P_{a},B_{b}\right] =-\epsilon _{ab}Y\,,\qquad     %
\left[ H,P_{a}\right] =\frac{1}{\ell^2}\epsilon _{ab}G_{b}\,,
\notag \\
\left[ H,T_{a}\right] &=&\frac{1}{\ell^2}\epsilon _{ab}B_{b}\,,\ \ \ \,  %
\left[ P_{a},P_{b}\right] =-\frac{1}{\ell^2}\epsilon _{ab}S\,,
\notag \\
\left[ M,P_{a}\right] &=&\frac{1}{\ell^2}\epsilon _{ab}B_{b}\,,\ \ \ \,  %
\left[ P_{a},T_{b}\right] =-\frac{1}{\ell^2}\epsilon _{ab}Z\,. \label{EN}
\end{eqnarray}
Let us note that $\ell$ is a length parameter related to the cosmological constant through $\Lambda \propto \pm \frac{1}{\ell^2}$. In particular, in the vanishing cosmological constant limit $\ell\rightarrow\infty$, the algebra corresponds to the extended Newtonian algebra introduced in \cite{Ozdemir:2019orp}.  One can notice that the extended Newton-Hooke algebra \cite{Papageorgiou:2010ud,Hartong:2016yrf} appears setting $B_a,\,T_a,\,Y$ and $Z$ to zero.

The exotic Newtonian algebra admits the extended Newton-Hooke non-vanishing components of the invariant tensor\cite{Papageorgiou:2010ud,Hartong:2016yrf}
\begin{eqnarray}
\langle J S \rangle&=&-\alpha_0\,,\notag \\ 
\langle G_a G_b \rangle&=&\alpha_0\delta_{ab}\,, \notag\\ 
\langle J M \rangle&=&\langle H S\rangle=-\alpha_1\,, \notag \\ 
\langle G_a P_b \rangle&=&\alpha_1 \delta_{ab}\,, \notag \\
\langle H M \rangle &=&-\frac{\alpha_0}{\ell^2}\,, \notag \\
\langle P_a P_b \rangle&=&\frac{\alpha_0}{\ell^2}\delta_{ab}\,, \label{invt1}
\end{eqnarray}
along with\cite{Concha:2019dqs}
\begin{eqnarray}
\langle S S \rangle&=&\langle J Z \rangle=-\beta_0\,, \notag \\
\langle G_a B_b \rangle&=&\beta_0\delta_{ab}\,, \notag \\
\langle M S \rangle&=&\langle H Z\rangle=\langle J Y \rangle=-\beta_1\,, \notag \\
\langle P_a B_b \rangle&=&\langle G_a T_b \rangle=\beta_1 \delta_{ab}\,, \notag \\
\langle M M \rangle &=& \langle H Y \rangle=-\frac{\beta_0}{\ell^2}\,, \notag \\
\langle P_a T_b \rangle&=&\frac{\beta_0}{\ell^2}\delta_{ab}\,.\label{invt2}
\end{eqnarray}
Here, $\alpha_0,\,\alpha_1,\,\beta_0$ and $\beta_1$ are arbitrary independent constants. In particular, the components of the invariant tensor proportional to $\alpha_0$ and $\beta_0$ are related to the ``exotic'' sector of the theory \cite{Witten:1988hc}. On the other hand, the flat limit $\ell\rightarrow\infty$ reproduces the invariant tensor of the extended Bargmann algebra \cite{Bergshoeff:2016lwr,Hartong:2016yrf} and the most general extended Newtonian one \cite{Concha:2019dqs}. Let us note that, at the level of the exotic Newtonian algebra, the invariant tensor \eqref{invt1} alone is degenerate. Nevertheless, one obtain a non-degenerate bilinear form considering both families of invariant tensor given by \eqref{invt1} and \eqref{invt2}.

The CS action based on the exotic Newtonian algebra can be obtained considering the following gauge connection one-form:
\begin{eqnarray}
A=\tau H+e^{a}P_{a}+\omega J+\omega^{a}G_a+mM+sS+t^{a}T_a+b^{a}B_a+yY+zZ\,,\
\end{eqnarray}
and the non-vanishing components of the invariant tensor \eqref{invt1} and \eqref{invt2} into the general expression of a three-dimensional CS action,
\begin{eqnarray}
I_{CS}=\frac{k}{4\pi}\int\langle AdA+\frac{2}{3}A^3\rangle\,.\label{CS}
\end{eqnarray}
Here $k$ is the CS level of the theory which, for gravitational theories, is related to the gravitational constant $G$ through $k=1/(4G)$. The three-dimensional exotic Newtonian CS gravity action reads, up to boundary terms, as follows \cite{Concha:2019dqs}:
\begin{eqnarray}
I_{\text{Exotic-N}}=\frac{k}{4\pi}\int \mathcal{L}_{\text{Extended-NH}} +\mathcal{L}_{\text{Enhanced-BNH}}\,,
\end{eqnarray}
where
\begin{eqnarray}
\mathcal{L}_{\text{Extended-NH}}&=&\alpha_0\left[\omega_aR^{a}\left(\omega^{b}\right)-2sR\left(\omega\right)+\frac{1}{\ell^2}e_aR^{a}\left(e^{b}\right)-\frac{2}{\ell^2}mR\left(\tau\right)\right] \notag \\
& & +\,\alpha_1\left[e_aR^{a}\left(\omega^{b}\right)+\omega_aR^{a}\left(e^{b}\right)-2mR\left(\omega\right)-2sR\left(\tau\right) \right]\,,\label{ENH}
\end{eqnarray}
and
\begin{eqnarray}
\mathcal{L}_{\text{Enhanced-BNH}}&=&\beta_0\left[b_aR^{a}\left(\omega^{b}\right)+\omega_aR^{a}\left(b^{b}\right) - 2zR\left(\omega\right)-sds-\frac{2}{\ell^2}yR\left(\tau\right)-\frac{1}{\ell^2}mdm \right. \notag \\
& & \left.+\frac{1}{\ell^2}t_aR^{a}\left(e^{b}\right)+\frac{1}{\ell^2}e_aR^{a}\left(t^{b}\right)\right]+\beta_1\left[e_aR^{a}\left(b^{b}\right)+b_aR^{a}\left(e^{b}\right) \right. \notag \\
& &\left. +t_aR^{a}\left(\omega^{b}\right)+\omega_aR^{a}\left(t^{b}\right)-2yR\left(\omega\right)-2zR\left(\tau\right)-2mds\right].\label{eBNH}
\end{eqnarray}
Here we have used the curvatures
\begin{eqnarray}
R\left(\omega\right)&=&d\omega\,, \notag \\
R\left(\tau\right)&=&d\tau\,, \notag \\
R^{a}\left(\omega^{b}\right)&=&d\omega^{a}+\epsilon^{ac}\omega\omega_c+\frac{1}{\ell^2}\epsilon^{ac}\tau e_c\,, \notag\\
R^{a}\left(e^{b}\right)&=&de^{a}+\epsilon^{ac}\omega e_c+\epsilon^{ac}\tau\omega_c\,, \notag \\
R^{a}\left(t^{b}\right)&=&dt^{a}+\epsilon^{ac}\omega t_c+\epsilon^{ac}\tau b_c+\epsilon^{ac}s e_c+\epsilon^{ac}m\omega_c\,, \notag \\
R^{a}\left(b^{b}\right)&=&db^{a}+\epsilon^{ac}\omega b_c+\epsilon^{ac}s\omega_c+\frac{1}{\ell^2}\epsilon^{ac}\tau t_c+\frac{1}{\ell^2}\epsilon^{ac}me_c\,.\label{curv1}
\end{eqnarray}
Moreover, one can also introduce the curvatures
\begin{eqnarray}
R\left(m\right)&=&dm + \epsilon^{ac} \omega_a e_c \,, \notag \\
R\left(s\right)&=&ds + \frac{1}{2} \epsilon^{ac} \omega_a \omega_c + \frac{1}{2\ell^2} \epsilon^{ac} e_a e_c \,, \notag \\
R\left(y\right)&=&dy + \epsilon^{ac} \omega_a t_c + \epsilon^{ac} e_a b_c \,, \notag \\
R\left(z\right)&=&dz + \epsilon^{ac} \omega_a b_c + \frac{1}{\ell^2} \epsilon^{ac} e_a t_c \,. \label{curv1a}
\end{eqnarray}
While the first Lagrangian \eqref{ENH} corresponds to the extended Newton-Hooke gravity Lagrangian \cite{Papageorgiou:2010ud,Hartong:2016yrf}, the second one \eqref{eBNH} coincides with the enhanced Bargmann-Newton-Hooke gravity Lagrangian \cite{Bergshoeff:2020fiz} plus an exotic sector proportional to $\beta_0$. The extended Newton-Hooke gravity can be seen as the non-relativistic limit of the $\text{AdS}\,\oplus\mathfrak{u}\left(1\right)^2$ gravity. On the other hand, the enhanced Bargmann-Newton-Hooke gravity theory appears as a non-relativistic limit of the coadjoint $\text{AdS}\,\oplus\mathfrak{u}\left(1\right)^2$ gravity \cite{Bergshoeff:2020fiz}. Both Lagrangians define the exotic Newtonian gravity theory which in the vanishing cosmological constant limit $\ell\rightarrow\infty$ reproduces the extended Newtonian gravity theory \cite{Ozdemir:2019orp,Concha:2019dqs}.
\section{Exotic Newtonian superalgebra and semigroup expansion method}\label{sec3}
In this section, we show that the exotic Newtonian superalgebra can be obtained by considering the $S$-expansion \cite{Izaurieta:2006zz} of the relativistic $\mathcal{N}=2$ AdS superalgebra. Our approach provides not only the (anti-)commutation relations of the exotic Newtonian superalgebra but also the non-vanishing components of the invariant tensor, which are essential for the construction of a CS action. Before presenting the explicit derivation of the exotic Newtonian superalgebra, we first review the $S$-expansion procedure and its main features.

The expansion method was first introduced in \cite{Hatsuda:2001pp} and subsequently developed in \cite{deAzcarraga:2002xi,Izaurieta:2006zz} to obtain a new (bigger) Lie algebra $\mathfrak{G}$ from an original one $\mathfrak{g}$. In particular, the expansion based on semigroups \cite{Izaurieta:2006zz, Caroca:2011qs, Andrianopoli:2013ooa, Artebani:2016gwh, Ipinza:2016bfc, Penafiel:2016ufo, Inostroza:2018gzd} allows to define a new Lie algebra, $\mathfrak{G}=S\times\mathfrak{g}$, by combining the structure constants of a Lie algebra $\mathfrak{g}$ with the elements of a semigroup $S$. The $S$-expanded Lie (super)algebra satisfies
\begin{eqnarray}
\left[T_{\left(A,\alpha\right)},T_{\left(B,\beta\right)}\right]&=&{K_{\alpha\beta}}^{\gamma}{C_{AB}}^{C} T_{\left(C,\gamma\right)}\,,
\end{eqnarray}
where ${C_{AB}}^{C}$ are the structure constants of the original Lie (super)algebra $\mathfrak{g}$, $T_{\left(A,\alpha\right)}=\lambda_\alpha T_A$ denote the generators of the expanded (super)algebra $\mathfrak{G}$, and ${K_{\alpha\beta}}^{\gamma}$ is the so-called 2-selector, which satisfies
\begin{equation}
{K_{\alpha\beta}}^{\gamma}=\left\{
\begin{array}{c}
1\text{\qquad when }\lambda _{\alpha }\lambda _{\beta }=\lambda _{\gamma } \,,
\\
0\qquad \text{otherwise. \ \ \ \ \ \ \ \ }%
\end{array}%
\right.
\end{equation}
A smaller (super)algebra can be extracted from the expanded one, called as $0_S$-reduced (super)algebra, by considering a semigroup with zero element $0_S\in S$ and imposing the condition $0_ST_A=0$. An alternative method to extract a smaller (super)algebra from the expanded one $\mathfrak{G}$ requires to consider a subset decomposition of the semigroup $S=\bigcup_{p\in I}S_p$ which satisfies the same structure than a subspace decomposition of the original (super)algebra $\mathfrak{g}=\bigoplus_{p\in I}V_p$, where $I$ denotes a set of indices. Then, the subalgebra
\begin{equation}
\mathfrak{G}_{R}=\bigoplus\limits_{p\in I}S_{p}\times V_{p}\,,
\end{equation}
is called the resonant subalgebra of the expanded (super)algebra $\mathfrak{G}$.

It is important to mention that, unlike the power series expansion carried out on the Maurer-Cartan forms \cite{deAzcarraga:2002xi}, the $S$-expansion method is defined entirely on the Lie (super)algebra $\mathfrak{g}$ without considering the group manifold. Furthermore, for a particular choice of the semigroup $S$, the $S$-expansion procedure can reproduce the expanded (super)algebras obtained through the Maurer-Cartan forms power series expansion. Moreover, the $S$-expansion method provides us with the non-vanishing components of the invariant tensor of the $S$-expanded (super)algebra in terms of the invariant tensor of the original (super)algebra. It is important to mention that both expansion mechanisms have been useful not only to construct new relativistic (super)gravity theories \cite{Izaurieta:2006aj, Edelstein:2006se, deAzcarraga:2007et, Izaurieta:2009hz, Concha:2013uhq, Salgado:2014jka, Concha:2014vka, Concha:2014tca, Durka:2016eun, Penafiel:2017wfr, Caroca:2017izc, Caroca:2017onr, Penafiel:2018vpe, Caroca:2019dds} but also at the non-relativistic level \cite{Bergshoeff:2019ctr, deAzcarraga:2019mdn, Romano:2019ulw, Fontanella:2020eje, Concha:2019lhn, Penafiel:2019czp, Gomis:2019nih, Kasikci:2020qsj, Concha:2020sjt, Concha:2020ebl, Concha:2020eam}.
\subsection{Exotic Newtonian superalgebra by expanding the \texorpdfstring{$\mathcal{N}=2$}{N2} AdS superalgebra}

Here we explore a supersymmetric extension of the exotic Newtonian algebra by considering the $S$-expansion of the $\mathcal{N}=2$ AdS superalgebra given by an $\mathfrak{so}\left(2\right)$ extension of the $\mathfrak{osp}\left(2,2\right)\otimes\mathfrak{sp}\left(2\right)$ superalgebra. To this end, we consider the procedure used in \cite{Gomis:2019nih,Concha:2020tqx} to obtain a proper exotic Newtonian superalgebra which admits an invariant supertrace.

Let us first consider an $\mathfrak{so}\left(2\right)$ extension of the $\mathfrak{osp}\left(2,2\right)\otimes\mathfrak{sp}\left(2\right)$ superalgebra which is spanned by the set of generators $\{\tilde{J}_A,\tilde{P}_A,\tilde{\mathcal{T}},\tilde{\mathcal{U}},\tilde{Q}_{\alpha}^{i}\}$ and satisfies the following (anti-)commutation relations \cite{Howe:1995zm}:
\begin{eqnarray}
\left[ \tilde{J}_{A},\tilde{J}_{B}\right] &=&\epsilon _{ABC}\tilde{J}^{C}\,, \notag \\ \left[ \tilde{J}_{A},\tilde{P}_{B}\right] &=&\epsilon _{ABC}\tilde{P}^{C}\,, \notag \\
\left[ \tilde{P}_{A},\tilde{P}_{B}\right] &=&\frac{1}{\ell^2}\epsilon _{ABC}\tilde{J}^{C}\,, \notag \\
\left[\tilde{J}_{A},\tilde{Q}_{\alpha}^{i}\right] &=& -\frac{1}{2} \left( \gamma_{A} \right)_{\alpha}^{\ \beta} \tilde{Q}_{\beta}^{i}\,, \notag \\ 
\left[\tilde{P}_{A},\tilde{Q}_{\alpha}^{i}\right] &=& -\frac{1}{2\ell} \left( \gamma_{A} \right)_{\alpha}^{\ \beta} \tilde{Q}_{\beta}^{i}\,, \notag \\
\left[\tilde{\mathcal{T}},\tilde{Q}_{\alpha}^{i}\right] &=& \frac{1}{2} \epsilon^{ij} \tilde{Q}_{\beta}^{j}\,, \notag \\
\{ \tilde{Q}_{\alpha}^{i},\tilde{Q}_{\beta}^{j}\} &=&-\frac{\delta_{ij}}{\ell}\left(\gamma^{A}C \right)_{\alpha \beta} \tilde{J}_{A}-\delta_{ij}\left(\gamma^{A}C \right)_{\alpha \beta} \tilde{P}_{A}-C_{\alpha \beta}\epsilon^{ij}\left(\tilde{\mathcal{U}}+\frac{1}{\ell}\tilde{\mathcal{T}}\right)\,, \label{n2SADS}
\end{eqnarray}%
where $\alpha,\beta=1,2$, $i,j=1,2$ denotes the number of supercharges and $A,B,C=0,1,2$ are the Lorentz indices which are raised and lowered with the Minkoswki metric $\eta_{AB}=(-1,1,1)$. Here, $C$ is the charge conjugation matrix,
\begin{equation}
C_{\alpha\beta}=C^{\alpha\beta}=\begin{pmatrix}
0 & -1\\
1 & 0\\
\end{pmatrix}\, ,
\end{equation}
which satisfies $C^{T}=-C$ and $C\gamma^{A}=(C\gamma^{A})^{T}$ with $\gamma^{A}$ being the Dirac matrices in three spacetime dimensions. Let us note that the presence of the generators $\{\tilde{\mathcal{T}},\tilde{\mathcal{U}}\}$ is required not only to construct an exotic Newtonian superalgebra with non-degenerate invariant tensor, but also to establish a proper flat limit $\ell\rightarrow\infty$. The non-degeneracy of the invariant tensor is crucial  to get
a well-defined CS (super)gravity action and is related to the Physical requirement that the CS action involves a kinematical term for each gauge field. In particular, the non-vanishing components of the invariant tensor for the $\mathfrak{so}\left(2\right)$ extension of the $\mathfrak{osp}\left(2,2\right)\otimes\mathfrak{sp}\left(2\right)$ superalgebra read
\begin{eqnarray}
\langle\tilde{J}_{A}\tilde{J}_{B}\rangle&=&\mu_0\eta_{AB}\,,\notag \\
\langle\tilde{J}_{A}\tilde{P}_{B}\rangle&=&\mu_1\eta_{AB}\,,\notag \\
\langle\tilde{P}_A \tilde{P}_B\rangle&=&\frac{\mu_0}{\ell^2}\eta_{AB}\,,\notag \\
\langle\tilde{\mathcal{T}}\tilde{\mathcal{T}}\rangle&=&\mu_0\,,\notag \\
\langle\tilde{\mathcal{T}}\tilde{\mathcal{U}}\rangle&=&\mu_1\,,\notag \\
\langle\tilde{\mathcal{U}}\tilde{\mathcal{U}}\rangle&=&-\frac{\mu_1}{\ell^2}\,,\notag \\
\langle \tilde{Q}^{i}_{\alpha}\tilde{Q}^{j}_{\beta}\rangle&=&2\left(\mu_1+\frac{\mu_0}{\ell}\right)C_{\alpha\beta}\delta^{ij}\,, \label{invt3}
\end{eqnarray}
where $\mu_0$ and $\mu_1$ are arbitrary constants. In the flat limit $\ell\rightarrow\infty$, the invariant tensor corresponds to the $\mathcal{N}=2$ super-Poincaré ones \cite{Howe:1995zm}.

Before considering the $S$-expansion of the $\mathcal{N}=2$ AdS superalgebra \eqref{n2SADS}, it is convenient to decompose the indices $A,B$ as
\begin{eqnarray}
A\rightarrow\left(0,a\right)
\end{eqnarray}
with $a=1,2$.
Then let $V_0=\{\tilde{J}_0,\tilde{P}_0,\tilde{\mathcal{T}},\tilde{\mathcal{U}},\tilde{Q}^{+}_{\alpha}\}$ and $V_1=\{\tilde{J}_a,\tilde{P}_a,\tilde{Q}^{-}_{\alpha}\}$ be a subset decomposition of the $\mathcal{N}=2$ AdS superalgebra where
\begin{eqnarray}
\tilde{Q}^{\pm}_{\alpha}=\frac{1}{\sqrt{2}}\left(\tilde{Q}^{1}_{\alpha}\pm\epsilon_{\alpha\beta}\tilde{Q}^{2}_{\beta}\right)\,.
\end{eqnarray}
One can see that the subspace decomposition of the $\mathcal{N}=2$ AdS superalgebra satisfies
\begin{eqnarray}
\left[V_0,V_0\right]\subset V_0\,, \qquad \left[V_0,V_1\right]\subset V_1\,, \qquad \left[V_1,V_1\right]\subset V_0\,. \label{subdec}
\end{eqnarray}

Let us consider now $S_{E}^{\left(4\right)}=\{\lambda_0,\lambda_1,\lambda_2,\lambda_3,\lambda_4,\lambda_5\}$ as the relevant abelian semigroup whose elements satisfy the following multiplication law:
\begin{equation}
\lambda _{\alpha }\lambda _{\beta }=\left\{
\begin{array}{lcl}
\lambda _{\alpha +\beta }\,\,\,\, & \mathrm{if}\,\,\,\,\alpha +\beta <5\,, &
\\
\lambda _{5}\,\, & \mathrm{if}\,\,\,\,\alpha +\beta \geq 5\,, &
\end{array}%
\right.  \label{ml}
\end{equation}
with $\lambda_5=0_S$ being the zero element of the semigroup which satisfies $0_S\lambda_{i}=0_S$. Then, let $S_{E}^{\left(4\right)}=S_0\cup S_1$ be a resonant semigroup decomposition where
\begin{eqnarray}
S_0&=&\{\lambda_0,\lambda_2,\lambda_4,\lambda_5\}\,,\\
S_1&=&\{\lambda_1,\lambda_3,\lambda_5\}\,.\label{semdec}
\end{eqnarray}
One can see that such semigroup decomposition satisfies the same algebraic structure than the subspace decomposition,
\begin{eqnarray}
S_0\cdot S_0 \subset S_0\,,\qquad S_0\cdot S_1 \subset S_1\,, \qquad S_1\cdot S_1\subset S_0\,. \label{resonant}
\end{eqnarray}
Then one finds a new expanded superalgebra after extracting a resonant subalgebra from the $S_{E}^{\left(4\right)}$-expansion of the $\mathcal{N}=2$ AdS superalgebra,
\begin{eqnarray}
\mathfrak{G}_{R}&=&S_{0}\times V_{0}\oplus S_{1}\times V_{1}\,,
\end{eqnarray}
and considering a $0_S$-reduction. The novel superalgebra is spanned by the exotic Newtonian bosonic generators
\begin{equation}
    \{J,G_{a},S,B_{a},Z,H,P_{a},M,T_{a},Y\} \label{set1}
\end{equation}
along with
\begin{equation}
    \{Y_1,Y_2,Y_3,U_1,U_2,U_3,Q^{+}_{\alpha},Q^{-}_{\alpha},R_{\alpha},W^{-}_{\alpha},W^{+}_{\alpha}\} \,. \label{set2}
\end{equation}
The expanded generators $\eqref{set1}$ and $\eqref{set2}$ are related to the $\mathcal{N}=2$ super AdS ones through
\begin{equation}
    \begin{tabular}{lll}
\multicolumn{1}{l|}{$\lambda_5$} & \multicolumn{1}{|l}{\cellcolor[gray]{0.8}} & \multicolumn{1}{|l|}{\cellcolor[gray]{0.8}} \\ \hline
\multicolumn{1}{l|}{$\lambda_4$} & \multicolumn{1}{|l}{$Z,\ Y,\ \,Y_3,\ U_3,\,W^{+}_{\alpha}$} & \multicolumn{1}{|l|}{\cellcolor[gray]{0.8}} \\ \hline
\multicolumn{1}{l|}{$\lambda_3$} & \multicolumn{1}{|l}{\cellcolor[gray]{0.8}} & \multicolumn{1}{|l|}{$B_a,\,T_a,\,W^{-}_{\alpha}$} \\ \hline
\multicolumn{1}{l|}{$\lambda_2$} & \multicolumn{1}{|l}{$S,\ M,\ Y_2,\,U_2,\,R_{\alpha}$} & \multicolumn{1}{|l|}{\cellcolor[gray]{0.8}} \\ \hline
\multicolumn{1}{l|}{$\lambda_1$} & \multicolumn{1}{|l}{\cellcolor[gray]{0.8}} & \multicolumn{1}{|l|}{$G_a,\,P_a,\,Q^{-}_{\alpha}$} \\ \hline
\multicolumn{1}{l|}{$\lambda_0$} & \multicolumn{1}{|l}{$ J,\,\ H,\ Y_1,\,U_1,\,Q^{+}_{\alpha}$} & \multicolumn{1}{|l|}{\cellcolor[gray]{0.8}} \\ \hline
\multicolumn{1}{l|}{} & \multicolumn{1}{|l}{$\tilde{J}_0,\,\tilde{P}_0,\,\tilde{\mathcal{T}},\ \tilde{\mathcal{U}},\ \, \tilde{Q}^{+}_{\alpha}$} & \multicolumn{1}{|l|}{$\tilde{J}_{a},\ \tilde{P}_{a},\,\tilde{Q}^{-}_{\alpha}$} 
\end{tabular}%
\end{equation}
Using the (anti-)commutation relations of the $\mathcal{N}=2$ AdS superalgebra \eqref{n2SADS} and the
multiplication law of the semigroup \eqref{ml}, one can show that the expanded generators satisfy the
exotic Newtonian algebra \eqref{EN} along with the following commutation relations:
\begin{eqnarray}
\left[ J,Q^{\pm}_{\alpha}\right] &=&-\frac{1}{2}\left(\gamma_0\right)_{\alpha}^{\ \beta}Q^{\pm}_{\beta}\,, \qquad \ %
\left[ J,R_{\alpha}\right] =-\frac{1}{2}\left(\gamma_0\right)_{\alpha}^{\ \beta}R_{\beta}\,, \qquad \ %
\left[ J,W^{\pm}_{\alpha}\right] =-\frac{1}{2}\left(\gamma_0\right)_{\alpha}^{\ \beta}W^{\pm}_{\beta}\,,  \notag
\\
\left[ S,Q^{+}_{\alpha}\right] &=&-\frac{1}{2}\left(\gamma_0\right)_{\alpha}^{\ \beta}R_{\beta}\,, \qquad \, %
\left[ S,Q^{-}_{\alpha}\right] =-\frac{1}{2}\left(\gamma_0\right)_{\alpha}^{\ \beta}W^{-}_{\beta}\,, \qquad \ %
\left[ S,R_{\alpha}\right] =-\frac{1}{2}\left(\gamma_0\right)_{\alpha}^{\ \beta}W^{+}_{\beta}\,,  \notag
\\
\left[ H,Q^{\pm}_{\alpha}\right] &=&-\frac{1}{2\ell}\left(\gamma_0\right)_{\alpha}^{\ \beta}Q^{\pm}_{\beta}\,, \quad \ \  %
\left[ H,R_{\alpha}\right] =-\frac{1}{2\ell}\left(\gamma_0\right)_{\alpha}^{\ \beta}R_{\beta}\,, \quad \ \, %
\left[ H,W^{\pm}_{\alpha}\right] =-\frac{1}{2\ell}\left(\gamma_0\right)_{\alpha}^{\ \beta}W^{\pm}_{\beta}\,,  \notag
\\
\left[ M,Q^{+}_{\alpha}\right] &=&-\frac{1}{2\ell}\left(\gamma_0\right)_{\alpha}^{\ \beta}R_{\beta}\,, \quad \   %
\left[ M,Q^{-}_{\alpha}\right] =-\frac{1}{2\ell}\left(\gamma_0\right)_{\alpha}^{\ \beta}W^{-}_{\beta}\,, \quad \ \, %
\left[ M,R_{\alpha}\right] =-\frac{1}{2\ell}\left(\gamma_0\right)_{\alpha}^{\ \beta}W^{+}_{\beta}\,,  \notag
\\
\left[ G_a,Q^{+}_{\alpha}\right] &=&-\frac{1}{2}\left(\gamma_a\right)_{\alpha}^{\ \beta}Q^{-}_{\beta}\,, \quad \ %
\left[ G_a,Q^{-}_{\alpha}\right] =-\frac{1}{2}\left(\gamma_a\right)_{\alpha}^{\ \beta}R_{\beta}\,, \qquad \, %
\left[ G_a,R_{\alpha}\right] =-\frac{1}{2}\left(\gamma_a\right)_{\alpha}^{\ \beta}W^{-}_{\beta}\,,  \notag
\\
\left[ G_a,W^{-}_{\alpha}\right] &=&-\frac{1}{2}\left(\gamma_a\right)_{\alpha}^{\ \beta}W^{+}_{\beta}\,, \quad \, %
\left[ B_a,Q^{\pm}_{\alpha}\right] =-\frac{1}{2}\left(\gamma_a\right)_{\alpha}^{\ \beta}W^{\mp}_{\beta}\,, \quad \, %
\left[ P_a,Q^{+}_{\alpha}\right] =-\frac{1}{2\ell}\left(\gamma_a\right)_{\alpha}^{\ \beta}Q^{-}_{\beta}\,,  \notag
\\
\left[ P_a,Q^{-}_{\alpha}\right] &=&-\frac{1}{2\ell}\left(\gamma_a\right)_{\alpha}^{\ \beta}R_{\beta}\,, \quad \ \  %
\left[ P_a,R_{\alpha}\right] =-\frac{1}{2\ell}\left(\gamma_a\right)_{\alpha}^{\ \beta}W^{-}_{\beta}\,, \ \  %
\left[ P_a,W^{-}_{\alpha}\right] =-\frac{1}{2\ell}\left(\gamma_a\right)_{\alpha}^{\ \beta}W^{+}_{\beta}\,,  \notag
\\
\left[ T_a,Q^{\pm}_{\alpha}\right] &=&-\frac{1}{2\ell}\left(\gamma_a\right)_{\alpha}^{\ \beta}W^{\mp}_{\beta}\,, \ \ \,  %
\left[ Y_1,Q^{\pm}_{\alpha}\right] =\pm\frac{1}{2}\left(\gamma_0\right)_{\alpha\beta}Q^{\pm}_{\beta}\,, \qquad  %
\left[ Y_1,R_{\alpha}\right] =\frac{1}{2}\left(\gamma_0\right)_{\alpha\beta}R_{\beta}\,,  \notag
\\
\left[ Y_1,W^{\pm}_{\alpha}\right] &=&\pm\frac{1}{2}\left(\gamma_0\right)_{\alpha\beta}W^{\pm}_{\beta}\,, \ \ \ \, %
\left[ Y_2,Q^{+}_{\alpha}\right] =\frac{1}{2}\left(\gamma_0\right)_{\alpha\beta}R_{\beta}\,, \qquad \ \,  %
\left[ Y_2,Q^{-}_{\alpha}\right] =-\frac{1}{2}\left(\gamma_0\right)_{\alpha\beta}W^{-}_{\beta}\,,  \notag
\\
\left[ Y_2,R_{\alpha}\right] &=&\frac{1}{2}\left(\gamma_0\right)_{\alpha\beta}W^{+}_{\beta}\,, \qquad  %
\left[ Y_3,Q^{+}_{\alpha}\right] =\frac{1}{2}\left(\gamma_0\right)_{\alpha\beta}W^{+}_{\beta}\,, \notag \\
\left[ Z,Q^+_{\alpha}\right]& =& -\frac{1}{2}\left(\gamma_0\right)_{\alpha}^{\ \beta}W^+_{\beta} \,, \quad \ \ \, \left[ Y,Q^+_{\alpha}\right] =-\frac{1}{2\ell}\left(\gamma_0\right)_{\alpha}^{\ \beta}W^+_{\beta} \,, \label{SEN}
\end{eqnarray}
and the following anti-commutation relations:
\begin{eqnarray}
\{Q^{+}_{\alpha},Q^{+}_{\beta}\}&=&-\frac{1}{\ell}\left(\gamma^0 C\right)_{\alpha\beta}J-\left(\gamma^0 C\right)_{\alpha\beta}H-\frac{1}{\ell}\left(\gamma^0 C\right)_{\alpha\beta} Y_1-\left(\gamma^0 C\right)_{\alpha\beta}U_1\,, \notag \\
\{Q^{+}_{\alpha},Q^{-}_{\beta}\}&=&-\frac{1}{\ell}\left(\gamma^{a} C\right)_{\alpha\beta}G_a-\left(\gamma^{a} C\right)_{\alpha\beta}P_a\,, \notag \\
\{Q^{-}_{\alpha},Q^{-}_{\beta}\}&=&-\frac{1}{\ell}\left(\gamma^0 C\right)_{\alpha\beta}S-\left(\gamma^0 C\right)_{\alpha\beta}M+\frac{1}{\ell}\left(\gamma^0 C\right)_{\alpha\beta} Y_2+\left(\gamma^0 C\right)_{\alpha\beta}U_2\,, \notag \\
\{Q^{+}_{\alpha},R_{\beta}\}&=&-\frac{1}{\ell}\left(\gamma^0 C\right)_{\alpha\beta}S-\left(\gamma^0 C\right)_{\alpha\beta}M-\frac{1}{\ell}\left(\gamma^0 C\right)_{\alpha\beta} Y_2-\left(\gamma^0 C\right)_{\alpha\beta}U_2\,, \notag \\
\{Q^{-}_{\alpha},R_{\beta}\}&=&-\frac{1}{\ell}\left(\gamma^{a} C\right)_{\alpha\beta}B_a-\left(\gamma^{a} C\right)_{\alpha\beta}T_a\,, \notag \\
\{Q^{+}_{\alpha},W^{-}_{\beta}\}&=&-\frac{1}{\ell}\left(\gamma^{a} C\right)_{\alpha\beta}B_a-\left(\gamma^{a} C\right)_{\alpha\beta}T_a\,, \notag \\
\{Q^{-}_{\alpha},W^{-}_{\beta}\}&=&-\frac{1}{\ell}\left(\gamma^0 C\right)_{\alpha\beta}Z-\left(\gamma^0 C\right)_{\alpha\beta}Y+\frac{1}{\ell}\left(\gamma^0 C\right)_{\alpha\beta} Y_3+\left(\gamma^0 C\right)_{\alpha\beta}U_3\,, \notag \\
\{Q^{+}_{\alpha},W^{+}_{\beta}\}&=&-\frac{1}{\ell}\left(\gamma^0 C\right)_{\alpha\beta}Z-\left(\gamma^0 C\right)_{\alpha\beta}Y-\frac{1}{\ell}\left(\gamma^0 C\right)_{\alpha\beta} Y_3-\left(\gamma^0 C\right)_{\alpha\beta}U_3\,, \notag \\
\{R_{\alpha},R_{\beta}\}&=&-\frac{1}{\ell}\left(\gamma^0 C\right)_{\alpha\beta}Z-\left(\gamma^0 C\right)_{\alpha\beta}Y-\frac{1}{\ell}\left(\gamma^0 C\right)_{\alpha\beta} Y_3-\left(\gamma^0 C\right)_{\alpha\beta}U_3\,. \label{SEN2}
\end{eqnarray}
The superalgebra given by $\eqref{EN}$, $\eqref{SEN}$ and $\eqref{SEN2}$ corresponds to a supersymmetric extension of the exotic Newtonin algebra \cite{Concha:2019dqs}. Let us note that the expansion procedure implies the presence of additional bosonic generators $\{Y_1,Y_2,Y_3,U_1,U_2,U_3\}$ which, as we shall see, ensure having not only a non-degenerate invariant tensor but also a well-defined non-vanishing cosmological constant limit $\ell\rightarrow\infty$. Indeed, in the flat limit, one can see that the superalgebra reduces to the extended Newtonian superalgebra introduced in \cite{Ozdemir:2019orp} in presence of the additional generators $\{Y_1,Y_2,Y_3\}$ and central charges $\{U_1,U_2,U_3\}$. The extra bosonic generators $\{Y_1,Y_2,Y_3\}$ are expansions of the relativistic $R$-symmetry generator $\mathcal{T}$ and act non-trivially on the fermionic charges.

Interestingly, the exotic Newtonian superalgebra $\eqref{EN}$, $\eqref{SEN}$ and $\eqref{SEN2}$ can be written as two copies of the so-called enhanced Nappi-Witten algebra \cite{Bergshoeff:2020fiz,Concha:2020eam}, one of which is augemented by supersymmetry. Indeed, let us consider the following redefinition of the generators:
\begin{eqnarray}
G_{a}&=&L_a-\tilde{L}_a\,, \qquad \qquad \ \  P_a=\frac{1}{\ell}\left(L_a+\tilde{L}_a \right)\,, \qquad \qquad \ \  Q_{\alpha}^{+}=\sqrt{\frac{2}{\ell}} \mathcal{Q}_{\alpha}^{+}\,,\notag \\
B_{a}&=&N_a-\tilde{N}_a\,, \qquad \qquad \ \,  T_a=\frac{1}{\ell}\left(N_a+\tilde{N}_a \right)\,, \qquad \qquad \  Q_{\alpha}^{-}=\sqrt{\frac{2}{\ell}} \mathcal{Q}_{\alpha}^{-}\,,\notag \\
J&=&L+\tilde{L}\,, \qquad \qquad \ \ \ \ \ H=\frac{1}{\ell}\left(L-\tilde{L} \right)\,, \qquad \qquad \ \ \ \ \ R_{\alpha}=\sqrt{\frac{2}{\ell}} \mathcal{R}_{\alpha}\,,\notag \\
S&=&N+\tilde{N}\,, \qquad \qquad \ \ \ \, M=\frac{1}{\ell}\left(N-\tilde{N} \right)\,, \qquad \qquad \ \   W_{\alpha}^{+}=\sqrt{\frac{2}{\ell}} \mathcal{W}_{\alpha}^{+} \,,\notag \\
Z&=&B+\tilde{B}\,, \qquad \qquad \ \ \ \ \, Y=\frac{1}{\ell}\left(B-\tilde{B} \right)\,, \qquad \qquad \ \ \   W_{\alpha}^{-}=\sqrt{\frac{2}{\ell}} \mathcal{W}_{\alpha}^{-} \,,\notag \\
Y_{1}&=&X_1-\tilde{X}_1\,, \qquad \qquad \ Y_{2}=X_2-\tilde{X}_2\,,  \qquad \qquad \qquad \ \ \ Y_{3}=X_3-\tilde{X}_3\,,\notag \\ 
U_1&=&\frac{1}{\ell}\tilde{X}_1\,, \qquad \qquad \quad \ \ \ U_2=\frac{1}{\ell}\tilde{X}_2\,, 
 \qquad \qquad \qquad \qquad \ \ U_3=\frac{1}{\ell}\tilde{X}_3 \,, \label{redef}
\end{eqnarray}
where the set of generators
\begin{eqnarray}
\{L,N,B,L_a,N_a,X_1,X_2,X_3,\mathcal{Q}^{+}_{\alpha},\mathcal{Q}^{-}_{\alpha},\mathcal{R}_{\alpha},\mathcal{W}^{-}_{\alpha},\mathcal{W}^{+}_{\alpha}\}
\end{eqnarray}
satisfies a supersymmetric extension of the enhanced Nappi-Witten algebra. In particular, the generators of the enhanced Nappi-Witten superalgebra obey the following commutation relations:
\begin{eqnarray}
\left[L,L_a\right]&=&\epsilon_{ab}L_b\,,\qquad \qquad \ \ \, \left[L_a,L_b\right]=-\epsilon_{ab}N\,, \qquad \qquad \ \, \left[N,L_a\right]=\epsilon_{ab}N_b\,,\notag \\
\left[L,N_a\right]&=&\epsilon_{ab}N_b\,,\qquad \qquad \ \ \left[L_a,N_b\right]=-\epsilon_{ab}B\,, \qquad \quad \ \ \ \, \left[L,\mathcal{Q}^{\pm}_{\alpha}\right]=-\frac{1}{2}\left(\gamma_0\right)_{\alpha}^{\ \beta}\mathcal{Q}^{\pm}_{\beta}\,,\notag \\
\left[L,\mathcal{R}_{\alpha}\right]&=&-\frac{1}{2}\left(\gamma_0\right)_{\alpha}^{\ \beta}\mathcal{R}_{\beta}\,,\quad \left[L,\mathcal{W}^{\pm}_{\alpha}\right]=-\frac{1}{2}\left(\gamma_0\right)_{\alpha}^{\ \beta}\mathcal{W}^{\pm}_{\beta}\,, \ \, \left[N,\mathcal{Q}^{+}_{\alpha}\right]=-\frac{1}{2}\left(\gamma_0\right)_{\alpha}^{\ \beta}\mathcal{R}_{\beta}\,,\notag \\
\left[N,\mathcal{Q}^{-}_{\alpha}\right]&=&-\frac{1}{2}\left(\gamma_0\right)_{\alpha}^{\ \beta}\mathcal{W}^{-}_{\beta}\,,\quad \left[N,\mathcal{R}_{\alpha}\right]=-\frac{1}{2}\left(\gamma_0\right)_{\alpha}^{\ \beta}\mathcal{W}^{+}_{\beta}\,, \ \ \left[B,\mathcal{Q}^{+}_{\alpha}\right]=-\frac{1}{2}\left(\gamma_0\right)_{\alpha}^{\ \beta}\mathcal{W}^{+}_{\beta}\,,\notag \\
\left[X_1,\mathcal{Q}^{\pm}_{\alpha}\right]&=&\pm\frac{1}{2}\left(\gamma_0\right)_{\alpha\beta}\mathcal{Q}^{\pm}_{\beta}\,,\ \, \, \left[X_1,\mathcal{R}_{\alpha}\right]=\frac{1}{2}\left(\gamma_0\right)_{\alpha\beta}\mathcal{R}_{\beta}\,, \ \ \  \left[X_1,\mathcal{W}^{\pm}_{\alpha}\right]=\pm\frac{1}{2}\left(\gamma_0\right)_{\alpha\beta}\mathcal{W}^{\pm}_{\beta}\,,\notag \\
\left[X_2,\mathcal{Q}^{+}_{\alpha}\right]&=&\frac{1}{2}\left(\gamma_0\right)_{\alpha\beta}\mathcal{R}_{\beta}\,,\quad  \,  \left[X_2,\mathcal{Q}^{-}_{\alpha}\right]=-\frac{1}{2}\left(\gamma_0\right)_{\alpha\beta}\mathcal{W}^{-}_{\beta}\,, \ \, \left[X_2,\mathcal{R}_{\alpha}\right]=\frac{1}{2}\left(\gamma_0\right)_{\alpha\beta}\mathcal{W}^{+}_{\beta}\,,\notag \\
\left[X_3,\mathcal{Q}^{+}_{\alpha}\right]&=&\frac{1}{2}\left(\gamma_0\right)_{\alpha\beta}\mathcal{W}^{+}_{\beta}\,,\quad  \left[L_a,\mathcal{Q}^{+}_{\alpha}\right]=-\frac{1}{2}\left(\gamma_a\right)_{\alpha}^{\ \beta}\mathcal{Q}^{-}_{\beta}\,, \ \ \left[L_a,\mathcal{Q}^{-}_{\alpha}\right]=-\frac{1}{2}\left(\gamma_a\right)_{\alpha}^{\ \beta}\mathcal{R}_{\beta}\,, \notag \\
\left[L_a,\mathcal{R}_{\alpha}\right]&=&-\frac{1}{2}\left(\gamma_a\right)_{\alpha}^{\ \beta}\mathcal{W}^{-}_{\beta}\,, \, \left[L_a,\mathcal{W}^{-}_{\alpha}\right]=-\frac{1}{2}\left(\gamma_a\right)_{\alpha}^{\ \beta}\mathcal{W}^{+}_{\beta}\,, \ \left[N_a,\mathcal{Q}^{+}_{\alpha}\right]=-\frac{1}{2}\left(\gamma_a\right)_{\alpha}^{\ \beta}\mathcal{W}^{-}_{\beta}\,, \, \notag \\
\left[N_a,\mathcal{Q}^{-}_{\alpha}\right]&=&-\frac{1}{2}\left(\gamma_a\right)_{\alpha}^{\ \beta}\mathcal{W}^{+}_{\beta}\,, \label{SENW1}
\end{eqnarray}
along with the following anti-commutation relations:
\begin{eqnarray}
\{\mathcal{Q}^{+}_{\alpha},\mathcal{Q}^{+}_{\beta}\}&=&-\left(\gamma_0 C\right)_{\alpha\beta}L-\left(\gamma_0 C\right)_{\alpha\beta} X_1\,, \quad \ \{\mathcal{Q}^{-}_{\alpha},\mathcal{Q}^{-}_{\beta}\}=-\left(\gamma_0 C\right)_{\alpha\beta}N+\left(\gamma_0 C\right)_{\alpha\beta} X_2\,, \notag \\
\{\mathcal{Q}^{+}_{\alpha},\mathcal{R}_{\beta}\}&=&-\left(\gamma_0 C\right)_{\alpha\beta}N-\left(\gamma_0 C\right)_{\alpha\beta} X_2\,, \quad \{\mathcal{Q}^{-}_{\alpha},\mathcal{W}^{-}_{\beta}\}=-\left(\gamma_0 C\right)_{\alpha\beta}B+\left(\gamma_0 C\right)_{\alpha\beta} X_3\,, \notag \\
\{\mathcal{Q}^{+}_{\alpha},\mathcal{W}^{+}_{\beta}\}&=&-\left(\gamma_0 C\right)_{\alpha\beta}B-\left(\gamma_0 C\right)_{\alpha\beta} X_3\,, \quad \ \,
\{\mathcal{R}_{\alpha},\mathcal{R}_{\beta}\}=-\left(\gamma_0 C\right)_{\alpha\beta}B-\left(\gamma_0 C\right)_{\alpha\beta} X_3\,,\notag \\
\{\mathcal{Q}^{+}_{\alpha},\mathcal{Q}^{-}_{\beta}\}&=&-\left(\gamma^{a} C\right)_{\alpha\beta}L_a\,, \qquad \qquad \qquad \ \ \ \,
\{\mathcal{Q}^{-}_{\alpha},\mathcal{R}_{\beta}\}=-\left(\gamma^{a} C\right)_{\alpha\beta}N_a\,, \notag \\
\{\mathcal{Q}^{+}_{\alpha},\mathcal{W}^{-}_{\beta}\}&=&-\left(\gamma^{a} C\right)_{\alpha\beta}N_a\,. \label{SENW2}
\end{eqnarray}
While the Nappi-Witten superalgebra \cite{Concha:2020tqx,Concha:2020eam} requires three fermionic generators $\{\mathcal{Q}^{+}_{\alpha},\mathcal{Q}^{-}_{\alpha},\mathcal{R}_{\alpha}\}$, the supersymmetric extension of the enhanced Nappi-Witten algebra requires the introduction of five fermionic generators $\{\mathcal{Q}^{+}_{\alpha},\mathcal{Q}^{-}_{\alpha},\mathcal{R}_{\alpha},\mathcal{W}^{+}_{\alpha},\mathcal{W}^{-}_{\alpha}\}$. The presence of additional Majorana fermionic charges is due to the extra bosonic content and such fermionic generators are necessary to satisfy the Jacobi identity.\footnote{Let us mention, here, that the presence of extra fermionic generators (namely besides the gravitino) in supergravity has been considered also in higher-dimensional (relativistic) cases, see e.g. \cite{Andrianopoli:2016osu, Andrianopoli:2017itj, Penafiel:2017wfr, Ravera:2018vra, Banaudi:2018zmh}.}  On the other hand, the set of generators
\begin{eqnarray}
\{\tilde{L},\tilde{N},\tilde{B},\tilde{L}_a,\tilde{N}_a,\tilde{X}_1,\tilde{X}_2,\tilde{X}_3\}
\end{eqnarray}
satisfies the bosonic enhanced Nappi-Witten algebra \cite{Bergshoeff:2020fiz,Concha:2020eam}, that is
\begin{eqnarray}
\left[\tilde{L},\tilde{L}_a\right]&=&\epsilon_{ab}\tilde{L}_b\,,\qquad \qquad \ \left[\tilde{L}_a,\tilde{L}_b\right]=-\epsilon_{ab}\tilde{N}\,, \qquad \qquad \ \left[\tilde{N},\tilde{L}_a\right]=\epsilon_{ab}\tilde{N}_b\,,\notag \\
\left[\tilde{L},\tilde{N}_a\right]&=&\epsilon_{ab}\tilde{N}_b\,,\qquad \qquad \,  \left[\tilde{L}_a,\tilde{N}_b\right]=-\epsilon_{ab}\tilde{B}\,,\label{ENWa} 
\end{eqnarray}
supplemented with the additional bosonic generators $\{\tilde{X}_1,\tilde{X}_2,\tilde{X}_3\}$. Although the extra bosonic generators $\{\tilde{X}_1,\tilde{X}_2,\tilde{X}_3\}$ do not appear explicitly into the commutation relations, they are essential to define a non-degenerate invariant tensor. It is important to mention that the exotic Newtonian superalgebra can alternatively be obtained by expanding the enhanced Nappi-Witten superalgebra \eqref{SENW1}-\eqref{SENW2}, following the same procedure used in \cite{Concha:2020eam} (more details about the expansion of the enhanced Nappi-Witten superalgebra and its invariant tensor are given in Appendix \ref{appa}).
\section{Exotic Newtonian supergravity action and flat limit}\label{sec4}
In this section, we shall explore the explicit construction of a CS action based on the exotic Newtonian superalgebra given by \eqref{EN}, \eqref{SEN} and \eqref{SEN2}. As we shall see, the exotic Newtonian superalgebra allows us to accommodate a cosmological constant to the Newtonian supergravity.

The non-vanishing components of the invariant tensor for the exotic Newtonian superalgebra can be obtained from the $\mathcal{N}=2$ super AdS ones considering Theorem VII of Ref. \cite{Izaurieta:2006zz}. In particular, the exotic Newtonian superalgebra admits the non-vanishing components of the invariant tensor for the extended Newton-Hooke superalgebra \cite{Concha:2020eam},
\begin{eqnarray}
\langle J S \rangle &=& -\alpha_0\,, \qquad \qquad \qquad   \langle G_a G_b \rangle = \alpha_0 \delta_{ab}\,,\notag \\
\langle J M \rangle &=& \langle H S \rangle =-\alpha_1\,, \qquad \ \,  \langle G_a P_b \rangle = \alpha_1 \delta_{ab}  \,, \notag \\
\langle H M \rangle &=& -\frac{\alpha_0}{\ell^2}\,, \qquad \qquad \qquad  \langle P_a P_b \rangle= \frac{\alpha_0}{\ell^2}\delta_{ab}\,, \notag \\
\langle Y_1 Y_2 \rangle &=&\alpha_0 \qquad \qquad \qquad \, \ \  \langle Q^{-}_{\alpha} Q^{-}_{\beta} \rangle = \langle Q^{+}_{\alpha} R_{\beta} \rangle = 2\left(\alpha_1+\frac{\alpha_0}{\ell}\right)C_{\alpha\beta}\,, \notag \\ 
\langle U_1 U_2 \rangle &=&-\frac{\alpha_1}{\ell^2} \,, \qquad \qquad \quad \ \ \ \langle Y_1 U_2 \rangle =\langle Y_2 U_1 \rangle=\alpha_1\,, \label{invt6}
\end{eqnarray}
along with
\begin{eqnarray}
\langle SS \rangle &=& \langle J Z \rangle =-\beta_0\,, \qquad \quad \ \ \, \ \ \langle G_a B_b \rangle =\beta_0 \delta_{ab}\,, \notag \\
\langle M S \rangle &=& \langle H Z \rangle = \langle J Y \rangle = -\beta_1\,, \ \ \,  \langle P_a B_b \rangle = \langle G_a T_b \rangle = \beta_1 \delta_{ab} \,, \notag \\
\langle M M \rangle &=& \langle H Y \rangle = -\frac{\beta_0}{\ell^2} \,, \qquad \quad \ \ \  \
\langle P_a T_b \rangle =\frac{\beta_0}{\ell^2}\delta_{ab} \,, \notag\\
\langle Y_1 Y_3 \rangle &=&\langle Y_2 Y_2 \rangle = \beta_0 \,, \qquad  \quad \ \, \ \langle Q^{+}_{\alpha} W^{+}_{\beta} \rangle = \langle Q^{-}_{\alpha}W^{-}_{\beta} \rangle = \langle R_{\alpha}R_{\beta}\rangle=2\left(\beta_1+\frac{\beta_0}{\ell}\right)C_{\alpha\beta} \,, \notag \\
\langle U_1 U_3 \rangle &=& \langle U_2 U_2 \rangle = -\frac{\beta_1}{\ell^2}\,, \qquad \quad \ \  \langle Y_1 U_3 \rangle = \langle Y_3 U_1 \rangle = \langle Y_2 U_2 \rangle = \beta_1\,,  \label{invt7}
\end{eqnarray}
where the extended Newtonian parameters are related to the relativistic $\mathcal{N}=2$ super AdS ones through the semigroup elements as
\begin{eqnarray}
\alpha_0&=&\lambda_2 \mu_0\,,\qquad \qquad \qquad \alpha_1=\lambda_2 \mu_1\,, \notag \\
\beta_0&=&\lambda_4 \mu_0\,, \qquad \qquad \qquad \beta_1=\lambda_4 \mu_1\,. \label{param}
\end{eqnarray}
One can notice that the invariant tensor is given by two families proportional to $\alpha$ and $\beta$, respectively. In particular, the components proportional to $\alpha_0$ and $\beta_0$ are related to an exotic sector of a non-relativistic supergravity theory. In the vanishing cosmological constant limit $\ell\rightarrow\infty$, we recover the invariant tensor for the extended Bargmann \cite{Bergshoeff:2016lwr} and the extended Newtonian superalgebras \cite{Ozdemir:2019orp}. Although one could consider additional components coming from the expansion with the $\lambda_0$ element, we shall omit such components since they are related to trivial CS terms. On the other hand, although the invariant tensor $\eqref{invt6}$ is non-degenerate for the extended Newton-Hooke superalgebra, the non-degeneracy of the invariant supertrace for the exotic Newtonian superalgebra requires also to consider the components given by \eqref{invt7}.  As we shall see, the non-degeneracy of the bilinear invariant supertrace implies that the CS action involves a kinematical term for each gauge field.

The gauge connection one-form for the exotic Newtonian superalgebra is given by
\begin{eqnarray}
A&=&\tau H+e^{a}P_{a}+\omega J+\omega^{a}G_a+mM+sS+t^{a}T_a+b^{a}B_a+yY+zZ+y_1Y_1+y_2Y_2+y_3Y_3\notag \\
& &+u_1U_1+u_2U_2+u_3U_3+\bar{\psi}^{+}Q^{+}+\bar{\psi}^{-}Q^{-}+\bar{\rho}R+\bar{\phi}^{+}W^{+}+\bar{\phi}^{-}W^{-}\,. \label{1f}
\end{eqnarray}
The corresponding curvature two-form $F=dA+\frac{1}{2}\left[A,A\right]$ reads
\begin{eqnarray}
F&=&R\left(\tau\right) H+R^{a}\left(e^{b}\right)P_{a}+R\left(\omega\right) J+R^{a}\left(\omega^{b}\right)G_a+R\left(m\right)M+R\left(s\right)S+R^{a}\left(t^{b}\right)T_a\notag \\ 
& &+F^{a}\left(b^{b}\right)B_a+F\left(y\right)Y+F\left(z\right)Z+F\left(y_1\right)Y_1+F\left(y_2\right)Y_2+F\left(y_3\right)Y_3+F\left(u_1\right)U_1\notag \\
& &+F\left(u_2\right)U_2+F\left(u_3\right)U_3+\nabla\bar{\psi}^{+}Q^{+}+\nabla\bar{\psi}^{-}Q^{-}+\nabla\bar{\rho}R+\nabla\bar{\phi}^{+}W^{+}+\nabla\bar{\phi}^{-}W^{-}\,.
\end{eqnarray}
Here, the bosonic curvature two-forms are given by
\begin{eqnarray}
F\left(\omega\right)&=&R\left(\omega\right)+\frac{1}{2\ell}\bar{\psi}^{+}\gamma^{0}\psi^{+}\,, \qquad \qquad \quad \ \ \ F\left(y_1\right)=dy_1+\frac{1}{2\ell}\bar{\psi}^{+}\gamma^{0}\psi^{+}\,, \notag \\
F^{a}\left(\omega^{b}\right)&=&R^{a}\left(\omega^{b}\right)+\frac{1}{\ell}\bar{\psi}^{+}\gamma^{a}\psi^{-}\,, \qquad \qquad \quad \, F\left(y_2\right)=dy_2-\frac{1}{2\ell}\bar{\psi}^{-}\gamma^{0}\psi^{-}+\frac{1}{\ell}\bar{\psi}^{+}\gamma^{0}\rho\,, \notag\\
F\left(\tau\right)&=&R\left(\tau\right)+\frac{1}{2}\bar{\psi}^{+}\gamma^{0}\psi^{+}\,, \qquad \qquad \quad \ \ \ \ \, F\left(y_3\right)=dy_3-\frac{1}{\ell}\bar{\psi}^{-}\gamma^{0}\phi^{-}+\frac{1}{\ell}\bar{\psi}^{+}\gamma^{0}\phi^{+}+\frac{1}{2\ell}\bar{\rho}\gamma^{0}\rho\,, \notag \\
F^{a}\left(e^{b}\right)&=&R^{a}\left(e^{b}\right)+\bar{\psi}^{+}\gamma^{a}\psi^{-}\,, \qquad \qquad \qquad  F\left(u_1\right)=du_1+\frac{1}{2}\bar{\psi}^{+}\gamma^{0}\psi^{+}\,, \notag \\
F\left(s\right)&=&R\left(s\right)+\frac{1}{2\ell}\bar{\psi}^{-}\gamma^{0}\psi^{-}+\frac{1}{\ell}\bar{\psi}^{+}\gamma^{0}\rho \,,\quad \ F\left(u_2\right)=du_2-\frac{1}{2}\bar{\psi}^{-}\gamma^{0}\psi^{-}+\frac{1}{2}\bar{\psi}^{+}\gamma^{0}\rho\,, \notag \\
F^{a}\left(b^{b}\right)&=&R^{a}\left(b^{b}\right)+\frac{1}{\ell}\bar{\psi}^{-}\gamma^{a}\rho+\frac{1}{\ell}\bar{\psi}^{+}\gamma^{a}\phi^{-}\,,  \ \ F\left(u_3\right)=du_3-\bar{\psi}^{-}\gamma^{0}\phi^{-}+\bar{\psi}^{+}\gamma^{0}\phi^{+}+\frac{1}{2}\bar{\rho}\gamma^{0}\rho\,, \notag \\
F\left(m\right)&=&R\left(m\right)+\frac{1}{2}\bar{\psi}^{-}\gamma^{0}\psi^{-}+\bar{\psi}^{+}\gamma^{0}\rho \,, \notag \\
F^{a}\left(t^{b}\right)&=&R^{a}\left(t^{b}\right)+\bar{\psi}^{-}\gamma^{a}\rho+\bar{\psi}^{+}\gamma^{a}\phi^{-}\,, \notag \\
F\left(z\right)&=&R\left(z\right)+\frac{1}{\ell}\bar{\psi}^{-}\gamma^0\phi^{-}+\frac{1}{\ell}\bar{\psi}^{+}\gamma^0\phi^{+}+\frac{1}{2\ell}\bar{\rho}\gamma^0\rho \,,    \notag \\
F\left(y\right)&=&R\left(y\right)+\bar{\psi}^{-}\gamma^0\phi^{-}+\bar{\psi}^{+}\gamma^0\phi^{+}+\frac{1}{2}\bar{\rho}\gamma^0\rho \,,  \label{curv2}
\end{eqnarray}
where $R\left(\omega\right)$, $R^{a}\left(\omega^{b}\right)$, $R\left(\tau\right)$, $R^{a}\left(e^{b}\right)$, $R\left(s\right)$, $R^{a}\left(b^b\right)$, $R\left(m\right)$, $R^{a}\left(t^{b}\right)$, $R\left(z\right)$ and $R\left(y\right)$ are the respective bosonic curvature two-forms \eqref{curv1}-\eqref{curv1a} of the exotic Newtonian algebra. On the other hand, the fermionic curvature two-forms reads
\begin{eqnarray}
\nabla \psi^{+}&=&d\psi^{+}+\frac{1}{2}\omega\gamma_0\psi^{+}+\frac{1}{2\ell}\tau\gamma_0\psi^{+}-\frac{1}{2}y_1\gamma_0\psi^{+}\,\notag \\
\nabla \psi^{-}&=&d\psi^{-}+\frac{1}{2}\omega\gamma_0\psi^{-}+\frac{1}{2\ell}\tau\gamma_0\psi^{-} +\frac{1}{2}\omega^{a}\gamma_a\psi^{+} +\frac{1}{2\ell}e^{a}\gamma_a\psi^{+}+\frac{1}{2}y_1\gamma_0\psi^{-}\,\notag \\
\nabla\rho&=&d\rho+\frac{1}{2}\omega\gamma_0\rho+\frac{1}{2}s\gamma_0\psi^{+}+\frac{1}{2}\omega^{a}\gamma_a\psi^{-}+\frac{1}{2\ell}\tau\gamma_0\rho+\frac{1}{2\ell}m\gamma_0\psi^{+}+\frac{1}{2\ell}e^{a}\gamma_a\psi^{-}\notag \\
& &-\frac{1}{2}y_1\gamma_0\rho-\frac{1}{2}y_2\gamma_0\psi^{+}\,,\notag \\
\nabla\phi^{+}&=&d\phi^{+}+\frac{1}{2}\omega\gamma_0\psi^{+}+\frac{1}{2}s\gamma_0\rho+\frac{1}{2}\omega^{a}\gamma_a\phi^{-}+\frac{1}{2}b^{a}\gamma_a\psi^{-}+\frac{1}{2\ell}\tau\gamma_0\phi^{+}+\frac{1}{2\ell}m\gamma_0\rho\notag \\
& &+\frac{1}{2\ell}e^{a}\gamma_a\phi^{-}+\frac{1}{2\ell}t^{a}\gamma_a\psi^{-}-\frac{1}{2}y_1\gamma_0\phi^{+}-\frac{1}{2}y_2\gamma_0\rho-\frac{1}{2}y_3\gamma_0\psi^{+}\,,\notag \\
\nabla \phi^{-} &=&d\phi^{-}+\frac{1}{2}\omega\gamma_0\phi^{-}+\frac{1}{2}s\gamma_0\psi^{-}+\frac{1}{2}\omega^{a}\gamma_a\rho+\frac{1}{2}b^{a}\gamma_a\psi^{+}+\frac{1}{2\ell}\tau\gamma_0\phi^{-}+\frac{1}{2\ell}m\gamma_0\psi^{-}\notag \\
& &+\frac{1}{2\ell}e^{a}\gamma_a\rho+\frac{1}{2\ell}t^{a}\gamma_a\psi^{+}+\frac{1}{2}y_1\gamma_0\phi^{-}+\frac{1}{2}y_2\gamma_0\psi^{-} +\frac{1}{2} z \gamma_0 \psi^+ + \frac{1}{2\ell} y \gamma_0 \psi^+ \,. \label{curv2a}
\end{eqnarray}
The gauge-invariant CS supergravity action based on the super exotic Newtonian algebra given by \eqref{EN}, \eqref{SEN} and \eqref{SEN2} can be obtained by combining the gauge connection one-form \eqref{1f} and the non-vanishing components of the invariant tensor \eqref{invt6} and \eqref{invt7} into the general expression for the CS action \eqref{CS}. In particular, the exotic Newtonian (EN) CS supergravity action can be written in terms of the super Extended Newton-Hooke (ENH) and super Enhanced Bargmann-Newton-Hooke (EBNH) supergravity Lagrangians, that is
\begin{eqnarray}
I_{\text{Super-EN}}=\frac{k}{4\pi}\int\mathcal{L}_{\text{Super ENH}}+\mathcal{L}_{\text{Super EBNH}}\,,\label{CSsEN}
\end{eqnarray}
where
\begin{eqnarray}
\mathcal{L}_{\text{Super-ENH}}&=&\mathcal{L}_{\text{Extended-NH}}+\alpha_0\left[2y_1dy_2-\frac{2}{\ell}\bar{\psi}^{-}\nabla\psi^{-}-\frac{2}{\ell}\bar{\psi}^{+}\nabla\rho-\frac{2}{\ell}\bar{\rho}\nabla\psi^{+}\right] \notag \\
& & +\,\alpha_1\left[2y_1du_2+2y_2du_1+\frac{2}{\ell^2}u_1du_2-2\bar{\psi}^{-}\nabla\psi^{-}-2\bar{\psi}^{+}\nabla\rho-2\bar{\rho}\nabla\psi^{+} \right]\,,\label{SENH}
\end{eqnarray}
and
\begin{eqnarray}
\mathcal{L}_{\text{Super-EBNH}}&=&\mathcal{L}_{\text{Enhanced-BNH}}+\beta_0\left[2y_1dy_3+y_2dy_2-\frac{2}{\ell}\bar{\psi}^+\nabla\phi^{+}-\frac{2}{\ell}\bar{\phi}^+\nabla\psi^{+}-\frac{2}{\ell}\bar{\psi}^{-}\nabla\phi^{-}\right.\notag \\
&&-\,\left.\frac{2}{\ell}\bar{\phi}^{-}\nabla\psi^{-}-\frac{2}{\ell}\bar{\rho}\nabla{\rho}\right]+\beta_1\left[2y_1du_3+2y_3du_1+2y_2du_2+\frac{2}{\ell^2}u_1du_3\right.\notag \\
&&+\,\left.\frac{1}{\ell^2}u_2du_2-2\bar{\psi}^+\nabla\phi^{+}-2\bar{\phi}^+\nabla\psi^{+}-2\bar{\psi}^{-}\nabla\phi^{-}-2\bar{\phi}^{-}\nabla\psi^{-}-2\bar{\rho}\nabla{\rho}\right]\,.\label{SEBNH}
\end{eqnarray}
Here, $\mathcal{L}_{\text{Extended-NH}}$ and $\mathcal{L}_{\text{Enhanced-BNH}}$ correspond to the Extended Newton-Hooke and Enhanced Bargmann-Newton-Hooke gravity Lagragians which are given by \eqref{ENH} and \eqref{eBNH}, respectively.  The exotic Newtonian supergravity CS action \eqref{CSsEN} generalizes the extended Newtonian supergravity theory \cite{Ozdemir:2019orp} by introducing a cosmological constant and considering six additional bosonic gauge fields $y_1$, $y_2$, $y_3$, $u_1$, $u_2$ and $u_3$. Interestingly, a supersymmetric extension of the extended Newtonian gravity \cite{Ozdemir:2019orp, Concha:2019dqs} appears in the vanishing cosmological constant limit $\ell\rightarrow\infty$. In such limit, the curvature two-forms reduce to the extended Newtonian ones. Furthermore, the fermionic gauge fields do not contribute anymore to the exotic sectors, along $\alpha_0$ and $\beta_0$, in the flat limit. Such exotic terms can be seen as the non-relativistic version of the exotic supergravity Lagrangian \cite{Witten:1988hc}. In particular, the exotic Newtonian supergravity CS action \eqref{CSsEN} can alternatively be recovered as an $S_E^{\left(4\right)}$-expansion of the relativistic $\mathcal{N}=2$ AdS supergravity CS action \cite{Achucarro:1989gm, Howe:1995zm, Giacomini:2006dr},
\begin{eqnarray}
I_{\text{AdS}}^{\mathcal{N}=2}=\int \mu_0\left[\omega_{A}d\omega^{A}+\frac{1}{3}\epsilon_{ABC}\omega^{A}\omega^{B}\omega^{C}+\frac{1}{\ell^2}e_A T^{A}+\mathtt{t}d\mathtt{t}-\frac{2}{\ell}\tilde{\bar{\psi}}^{i}\nabla\tilde{\psi}^{i}\right]\notag \\
+\mu_1\left[2e_AR^{A}+\frac{1}{3\ell^2}\epsilon_{ABC}e^{A}e^{B}e^{C}+\mathtt{t}d\mathtt{u}-\frac{1}{\ell^2}\mathtt{u}d\mathtt{u}-2\tilde{\bar{\psi}}^{i}\nabla\tilde{\psi}^{i}\right]\,,\label{AdSCS}
\end{eqnarray}
where
\begin{eqnarray}
R^{A}&=&d\omega^{A}+\frac{1}{2}\epsilon^{ABC}\omega_B\omega_C\,,\notag \\
T^{A}&=&de^{A}+\frac{1}{2}\epsilon^{ABC}\omega_{B}e_{C}\,,\notag \\
\nabla\tilde{\psi}^{i}&=&d\tilde{\psi}^{i}+\frac{1}{2}\omega^{A}\gamma_A\tilde{\psi}^{i}+\frac{1}{2\ell}e^{A}\gamma_{A}\tilde{\psi}^{i}+\mathtt{t}\epsilon^{ij}\tilde{\psi}^{j}\,.\label{curv3}
\end{eqnarray}
Indeed, the exotic Newtonian CS supergravity action appears after considering the $S_E^{\left(4\right)}$-expansion of the $\mathcal{N}=2$ super AdS gauge fields as
\begin{eqnarray}
\omega&=&\lambda_0\omega_0\,,\qquad \ \, s=\lambda_2\omega_0\,,\quad \ \ \, z=\lambda_4\omega_0\,,\quad \ \ \omega_a=\lambda_1\omega_a\,,\quad \ \ b_a=\lambda_3\omega_{a}\,,\notag \\
\tau&=&\lambda_0 e_0\,,\qquad \, m=\lambda_2 e_0\,,\qquad y=\lambda_4 e_0\,,\qquad e_a=\lambda_1 e_a\,, \qquad t_a=\lambda_3 e_a\,,\notag \\
\psi^{+}&=&\lambda_0\tilde{\psi}^{+}\,,\quad \ \ \ \rho=\lambda_2\tilde{\psi}^{+}\,,\ \ \   \phi^{+}=\lambda_4\tilde{\psi}^{+}\,, \quad \, \psi^{-}=\lambda_1\tilde{\psi}^{-}\,, \quad \, \phi^{-}=\lambda_3\tilde{\psi}^{-}\,,\notag \\
y_1&=&\lambda_0 \mathtt{t}\,,\qquad \   y_2=\lambda_2 \mathtt{t}\,, \qquad \,  y_3=\lambda_4 \mathtt{t}\,,\notag \\
u_1&=&\lambda_0 \mathtt{u}\,,\qquad \   u_2=\lambda_2 \mathtt{u}\,, \qquad  u_3=\lambda_4 \mathtt{u}\,,
\end{eqnarray}
along with the expanded parameters \eqref{param} and defining
\begin{equation}
    \tilde{\psi}_{\alpha}^{\pm}=\frac{1}{\sqrt{2}}\left( \tilde{\psi}_{\alpha}^{1}\pm \epsilon_{\alpha\beta}\tilde{\psi}_{\beta}^{2}\right)\,.\label{redeff2}
\end{equation}

On the other hand, the non-degeneracy of the invariant tensor \eqref{invt6} and \eqref{invt7} implies that the field equations derived from the CS action \eqref{CSsEN} are given by the vanishing of the curvature two-forms \eqref{curv2}-\eqref{curv2a}. Such curvatures transform covariantly under the supersymmetry transformation laws (the explicit supersymmetry transformation laws of the exotic Newtonian superalgebra are given in Appendix \ref{appab}).



\section{Conclusions}\label{sec5}

In this work we have presented an exotic Newtonian CS supergravity theory in presence of a cosmological constant. The underlying non-relativistic superalgebra, which we have called as exotic Newtonian superalgebra, is obtained as an $S$-expansion of the $\mathcal{N}=2$ AdS superalgebra considering a particular semigroup. Interestingly, we showed that the exotic Newtonian superalgebra can be written as two copies of the so-called enhanced Nappi-Witten algebra \cite{Bergshoeff:2020fiz, Concha:2020ebl}, one of which is supersymmetric. The new Newtonian supergravity action contains the extended Newton-Hooke CS supergravity term \cite{Ozdemir:2019tby} as a sub-case. Furthermore, we showed that the most general extended Newtonian supergravity action appears in the flat limit $\ell\rightarrow\infty$ which contains the extended Bargmann supergravity \cite{Bergshoeff:2016lwr} as a particular sub-case. It would be interesting to explore the matter coupling of the exotic Newtonian supergravity theory presented here. Although the exotic Newtonian supergravity theory differs from the extended Newtonian one at the action level, one could expect that the matter couplings of both theories behave similarly. 

It would be worth considering further extensions of the Newtonian supergravity theory obtained here and in \cite{Ozdemir:2019orp} by considering a Maxwellian generalization. One could expect that such generalizations are given by the respective supersymmetric extensions of the enlarged and Maxwellian extended Newtonian gravity presented in \cite{Concha:2020ebl}.  On the other hand, following \cite{Ozdemir:2019tby}, one could explore a Schrödinger extension \cite{Bergshoeff:2015ija, Afshar:2015aku} of the exotic Newtonian superalgebra. Moving towards Newtonian supergravity theories and their Schrödinger extension can be useful, for instance, to approach supersymmetric field theories on non-relativistic curved backgrounds via localization \cite{Festuccia:2011ws,Pestun:2007rz,Marino:2012zq}.

Our results and those obtained in \cite{Gomis:2019nih, Concha:2020tqx} could be extended to other relativistic (super)algebras. Indeed, it seems that the $S_{E}^{\left(4\right)}$ semigroup allows to obtain the respective Newtonian version of a relativistic (super)algebra. In particular, the procedure used here could be useful in presence of supersymmetry, where the study of the non-relativistic limit is highly non-trivial. It is interesting to notice that the exotic Newtonian superalgebra can alternatively be recovered by expanding the enhanced Nappi-Witten superalgebra (see appendix \ref{appa}). Although both methods are based on the semigroup expansion method \cite{Izaurieta:2006zz}, they present subtle differences which could lead to diverse extensions of our results. Indeed, to obtain diverse Newtonian (super)algebras from an enhanced Nappi-Witten (super)algebra, we need to consider diverse semigroups as in \cite{Concha:2020ebl}. On the other hand, the derivation of various Newtonian (super)algebras by expanding a relativistic (super)algebra requires to consider different original algebras without modifying the semigroup.

\section*{Acknowledgments}

This work was funded by the National Agency for Research and Development ANID (ex-CONICYT) - PAI grant No. 77190078 (P.C.). This work was supported by the Research project Code DIREG$\_$09/2020 (P.C.) of the Universidad Católica de la Santisima Concepción, Chile. P.C. would like to thank to the Dirección de Investigación and Vice-rectoría de Investigación of the Universidad Católica de la Santísima Concepción, Chile, for their constant support.
L.R. would like to thank the Department of Applied Science and Technology of the Polytechnic University of Turin, and in particular Laura Andrianopoli and Francesco Raffa, for financial support.


\appendix

\section{Exotic Newtonian superalgebra by expanding an enhanced Nappi-Witten superalgebra}\label{appa}
As was shown in \cite{Concha:2020ebl}, diverse generalizations of the extended Newtonian algebra appear as $S$-expansions of an enhanced Nappi-Witten algebra. Here, we show that a supersymmetric extension of the enhanced Nappi-Witten algebra allows to obtain the exotic Newtonian superalgebra \eqref{EN}, \eqref{SEN} and \eqref{SEN2} considering a particular semigroup. Let
\begin{eqnarray}
\mathfrak{g}=\{\hat{L},\hat{N},\hat{B},\hat{L}_a,\hat{N}_a,\hat{X}_1,\hat{X}_2,\hat{X}_3,\hat{Q}^{+}_{\alpha},\hat{Q}^{-}_{\alpha},\hat{R}_{\alpha},\hat{W}^{-}_{\alpha},\hat{W}^{+}_{\alpha}\}
\end{eqnarray} 
be a supersymmetric extension of the enhanced Nappi-Witten algebra whose generators satisfy \eqref{SENW1} and \eqref{SENW2}. The presence of the additional generators $\{\hat{X}_1,\hat{X}_2,\hat{X}_3\}$ allows us not only to make contact with the exotic Newtonian superalgebra obtained here but also to establish a non-degenerate invariant tensor. In particular, the enhanced Nappi-Witten superalgebra \eqref{SENW1}-\eqref{SENW2} admits the following non-vanishing components of the invariant tensor:
\begin{eqnarray}
\langle \hat{L} \hat{N} \rangle &=&-\gamma_1\,,\qquad \qquad \qquad \qquad \quad \langle \hat{N} \hat{N} \rangle =\langle \hat{N} \hat{B} \rangle=-\gamma_2\,, \notag \\
\langle \hat{L}_a \hat{L}_b \rangle &=&\gamma_1 \delta_{ab}\,, \qquad \qquad \qquad \quad \quad
\langle \hat{L}_a \hat{N}_b \rangle =\gamma_2 \delta_{ab}\,, \notag\\
\langle \hat{X}_1 \hat{X}_2 \rangle &=&\varrho_1\,, \qquad \qquad \qquad \qquad \ \ \ \langle \hat{X}_2 \hat{X}_2 \rangle =\langle \hat{X}_1 \hat{X}_3 \rangle=\varrho_2\,, \notag \\ 
\langle \hat{Q}^{-}_{\alpha} \hat{Q}^{-}_{\beta}\rangle&=&\langle \hat{Q}^{+}_{\alpha} \hat{R}_{\beta}\rangle=2\gamma_1 C_{\alpha\beta}\,, \quad \ 
\langle \hat{Q}^{+}_{\alpha} \hat{W}^{+}_{\beta}\rangle=\langle \hat{Q}^{-}_{\alpha} \hat{W}^{-}_{\beta}\rangle=\langle \hat{R}_{\alpha}\hat{R}_{\beta}\rangle=2\gamma_2 C_{\alpha\beta}\,, \label{invt4}
\end{eqnarray}
where $\gamma_1,\, \varrho_1,\, \varrho_2$ and $\gamma_2$ are arbitrary independent constants. Let us note that the components proportional to $\gamma_1$ and $\rho_1$ correspond to those of the super Nappi-Witten ones \cite{Concha:2020eam}. Nevertheless, the complete set of components \eqref{invt4} is required to ensure the non-degeneracy of the bilinear invariant trace.

Before considering the $S$-expansion of the enhanced Nappi-Witten superalgebra, it is convenient to consider a subspace decomposition $\mathfrak{g}=V_0\oplus V_1$ where $V_0=\{\hat{L},\hat{N},\hat{B},\hat{L}_a,\hat{N}_a,\hat{X}_1,\hat{X}_2,\hat{X}_3\}$ and $V_1=\{\hat{Q}^{+}_{\alpha},\hat{Q}^{-}_{\alpha},\hat{R}_{\alpha},\hat{W}^{-}_{\alpha},\hat{W}^{+}_{\alpha}\}$. Such subspace decomposition satisfies \eqref{subdec}. On the other hand, let $S_{L}^{\left(1\right)}=\{\lambda_0,\lambda_1,\lambda_2\}$ be the relevant semigroup whose elements satisfy the following multiplication law \cite{Caroca:2019dds}:
\begin{equation}
\begin{tabular}{l|lll}
$\lambda _{2}$ & $\lambda _{2}$ & $\lambda _{2}$ & $\lambda _{2}$ \\
$\lambda _{1}$ & $\lambda _{2}$ & $\lambda _{1}$ & $\lambda _{2}$ \\
$\lambda _{0}$ & $\lambda _{0}$ & $\lambda _{2}$ & $\lambda _{2}$ \\ \hline
& $\lambda _{0}$ & $\lambda _{1}$ & $\lambda _{2}$%
\end{tabular}
\label{ml2}
\end{equation}%
where $\lambda_2=0_S$ is the zero element of the semigroup. Then, let us consider a semigroup decomposition $S_{L}^{\left(1\right)}=S_0\cup S_1$ with $S_0=\{\lambda_0,\lambda_1,\lambda_2\}$ and $S_1=\{\lambda_1,\lambda_2\}$,
which is said to be resonant since it satisfies the same algebraic structure as the subspace decomposition \eqref{subdec}. After considering a resonant subalgebra of the $S_{L}^{\left(1\right)}$-expansion of the enhanced Nappi-Witten superalgebra and performing a $0_S$-reduction, one finds an expanded superalgebra which is spanned by the set of generators
\begin{eqnarray}
\{L,N,B,L_a,N_a,X_1,X_2,X_3,\tilde{L},\tilde{N},\tilde{B},\tilde{L}_a,\tilde{N}_a,\tilde{X}_1,\tilde{X}_2,\tilde{X}_3,\mathcal{Q}^{+}_{\alpha},\mathcal{Q}^{-}_{\alpha},\mathcal{R}_{\alpha},\mathcal{W}^{-}_{\alpha},\mathcal{W}^{+}_{\alpha}\} \,.
\end{eqnarray}
The expanded generators are related to the original ones through the semigroup elements as
\begin{equation}
    \begin{tabular}{lll}
\multicolumn{1}{l|}{$\lambda_2$} & \multicolumn{1}{|l}{\cellcolor[gray]{0.8}} & \multicolumn{1}{|l|}{\cellcolor[gray]{0.8}} \\ \hline
\multicolumn{1}{l|}{$\lambda_1$} & \multicolumn{1}{|l}{$ L,\, N,\ B,\ L_a,\, N_a,\ X_1,\ X_2,\ X_3$} & \multicolumn{1}{|l|}{$\mathcal{Q}^{+}_{\alpha},\,\mathcal{Q}^{-}_{\alpha},\,\mathcal{R}_{\alpha},\,\mathcal{W}^{+}_{\alpha},\,\mathcal{W}^{-}_{\alpha}$} \\ \hline
\multicolumn{1}{l|}{$\lambda_0$} & \multicolumn{1}{|l}{$ \tilde{L},\, \tilde{N},\ \tilde{B},\ \tilde{L}_a,\, \tilde{N}_a,\ \tilde{X}_1,\ \tilde{X}_2,\ \tilde{X}_3$} & \multicolumn{1}{|l|}{\cellcolor[gray]{0.8}} \\ \hline
\multicolumn{1}{l|}{} & \multicolumn{1}{|l}{$\hat{L},\,\hat{N},\ \hat{B},\ \hat{L}_a,\,  \hat{N}_a, \ \hat{X}_1,\ \hat{X}_2,\  \hat{X}_3$} & \multicolumn{1}{|l|}{$\hat{Q}^{+}_{\alpha},\ \hat{Q}^{-}_{\alpha},\,\hat{R}_{\alpha},\,\hat{W}^{+}_{\alpha},\,\hat{W}^{-}_{\alpha}$} 
\end{tabular}%
\end{equation}
Two enhanced Nappi-Witten algebras, one of which is augmented by supersymmetry, are obtained using the multiplication law of the semigroup \eqref{ml2} and the (anti-)commutation relations of the original superalgebra \eqref{SENW1}-\eqref{SENW2}. In particular, the set of generators
\begin{eqnarray}
\{L,N,B,L_a,N_a,X_1,X_2,X_3,\mathcal{Q}^{+}_{\alpha},\mathcal{Q}^{-}_{\alpha},\mathcal{R}_{\alpha},\mathcal{W}^{-}_{\alpha},\mathcal{W}^{+}_{\alpha}\}
\end{eqnarray}
satisfies the enhanced Nappi-Witten superalgebra \eqref{SENW1}-\eqref{SENW2}. On the other hand, the set of generators
\begin{eqnarray}
\{\tilde{L},\tilde{N},\tilde{B},\tilde{L}_a,\tilde{N}_a,\tilde{X}_1,\tilde{X}_2,\tilde{X}_3\}
\end{eqnarray}
satisfies the bosonic enhanced Nappi-Witten algebra \eqref{ENWa}. Then, the exotic Newtonian superalgebra \eqref{EN}, \eqref{SEN} and \eqref{SEN2} appears by considering the redefinition of the generators as in \eqref{redef}. Let us note that the non-vanishing components of the invariant tensor for the two copies of the enhanced Nappi-Witten algebra, one of which is supersymmetric, can be obtained following the definitions of \cite{Izaurieta:2006zz}. Indeed, the expanded superalgebra admits the following non-vanishing components of the invariant tensor:
\begin{eqnarray}
\langle L_a L_b \rangle &=&\mu_1\delta_{ab}\,,\qquad \qquad \qquad \quad \ \ \langle \tilde{L}_a \tilde{L}_b \rangle =\nu_1\delta_{ab} \,, \notag \\
\langle L N \rangle &=&-\mu_1\,,\qquad \qquad \qquad \quad\ \ \ \ \ \langle \tilde{L} \tilde{N} \rangle =-\nu_1 \,, \notag \\
\langle X_1 X_2 \rangle &=&\rho_1\,,\qquad \qquad \qquad \quad \ \ \ \ \, \langle \tilde{X}_1 \tilde{X}_2 \rangle =\tilde{\rho}_1\,, \notag \\ 
\langle L_a N_b \rangle &=&\mu_2\delta_{ab}\,,\qquad \qquad \qquad \quad \ \langle \tilde{L}_a \tilde{N}_b \rangle =\nu_2\delta_{ab} \,, \notag \\
\langle N N \rangle &=&\langle N B \rangle = -\mu_2\,,\qquad \quad \ \ \ \ \,  \langle \tilde{N} \tilde{N} \rangle = \langle \tilde{N} \tilde{B} \rangle =-\nu_2 \, , \notag \\
\langle X_2 X_2 \rangle &=&\langle X_1 X_3 \rangle=\rho_2\,,\qquad  \quad \  \, \ \langle \tilde{X}_2 \tilde{X}_2 \rangle =\langle \tilde{X}_1 \tilde{X}_3 \rangle=\tilde{\rho}_2\,, \notag \\ 
\langle \mathcal{Q}^{-}_{\alpha} \mathcal{Q}^{-}_{\beta}\rangle&=&\langle \mathcal{Q}^{+}_{\alpha} \mathcal{R}_{\beta}\rangle=2\mu_1 C_{\alpha\beta}\,,\ \,   \langle \mathcal{Q}^{+}_{\alpha} \mathcal{W}^{+}_{\beta}\rangle=\langle \mathcal{Q}^{-}_{\alpha} \mathcal{W}^{-}_{\beta}\rangle=\langle \mathcal{R}_{\alpha}\mathcal{R}_{\beta}\rangle=2\mu_2 C_{\alpha\beta}\, , \label{invt5}
\end{eqnarray}
where the constants are related to the enhanced super Nappi-Witten ones through the semigroup elements as
\begin{eqnarray}
\mu_1&=&\lambda_1 \gamma_1\,, \qquad \mu_2=\lambda_1 \gamma_2\,, \qquad \rho_1=\lambda_1 \varrho_1\,, \qquad \rho_2=\lambda_1 \varrho_2\,, \notag \\
\nu_1&=&\lambda_0 \gamma_1\,, \qquad \nu_2=\lambda_0 \gamma_2\,, \qquad \tilde{\rho}_1=\lambda_0 \varrho_1\,, \qquad \tilde{\rho}_2=\lambda_0 \varrho_2\,.
\end{eqnarray}
Interestingly, the non-vanishing components of the invariant tensor for the exotic Newtonian superalgebra \eqref{invt6} and \eqref{invt7} can be alternatively recovered from the invariant tensor \eqref{invt5}, considering the redefinition of the generators as in \eqref{redef} and setting
\begin{eqnarray}
\alpha_0&=&\mu_1+\nu_1\,, \qquad \alpha_1=\frac{1}{\ell}\left(\mu_1-\nu_1\right)\,, \qquad \rho_1=\alpha_0+\alpha_1\,,\qquad \tilde{\rho}_1=-\alpha_{1}\,,\notag \\
\beta_0&=&\mu_2+\nu_2\,, \qquad \beta_1=\frac{1}{\ell}\left(\mu_2-\nu_2\right)\,, \qquad \rho_2=\beta_0+\beta_1\,,\qquad \, \tilde{\rho}_2=-\beta_{1}\,.
\end{eqnarray}


\section{Supersymmetry gauge transformations}\label{appab}
The supersymmetry transformation laws under which the curvature two-forms \eqref{curv2} and \eqref{curv2a} of the exotic Newtonian superalgebra transform in a covariant way read
\begin{eqnarray}
\delta\omega&=&-\frac{1}{\ell}\bar{\epsilon}^{+}\gamma^{0}\psi^{+}\,,\notag \\
\delta\omega^{a}&=&-\frac{1}{\ell}\bar{\epsilon}^{+}\gamma^{a}\psi^{-}-\frac{1}{\ell}\bar{\epsilon}^{-}\gamma^{a}\psi^{+}\,,\notag \\
\delta \tau&=&-\bar{\epsilon}^{+}\gamma^{0}\psi^{+}\,,\notag \\
\delta e^{a}&=&-\bar{\epsilon}^{+}\gamma^{a}\psi^{-}-\bar{\epsilon}^{-}\gamma^{a}\psi^{+}\,,\notag \\
\delta s&=&-\frac{1}{\ell}\bar{\epsilon}^{-}\gamma^{0}\psi^{-}-\frac{1}{\ell}\bar{\epsilon}^{+}\gamma^{0}\rho-\frac{1}{\ell}\bar{\eta}\gamma^{0}\psi^{+}\,,\notag \\
\delta m&=&-\bar{\epsilon}^{-}\gamma^{0}\psi^{-}-\bar{\epsilon}^{+}\gamma^{0}\rho-\bar{\rho}\gamma^{0}\psi^{+}\,,\notag \\
\delta b^{a}&=&-\frac{1}{\ell}\bar{\epsilon}^{-}\gamma^{a}\rho-\frac{1}{\ell}\bar{\eta}\gamma^{a}\psi^{-}-\frac{1}{\ell}\bar{\epsilon}^{+}\gamma^{a}\phi^{-}-\frac{1}{\ell}\bar{\zeta}^{-}\gamma^{a}\psi^{+}\,,\notag \\
\delta t^{a}&=&-\bar{\epsilon}^{-}\gamma^{a}\rho-\bar{\eta}\gamma^{a}\psi^{-}-\bar{\epsilon}^{+}\gamma^{a}\phi^{-}-\bar{\zeta}^{-}\gamma^{a}\psi^{+}\,,\notag \\
\delta z&=&-\frac{1}{\ell}\bar{\epsilon}^{-}\gamma^{0}\phi^{-}-\frac{1}{\ell}\bar{\zeta}^{-}\gamma^{0}\psi^{-}-\frac{1}{\ell}\bar{\epsilon}^{+}\gamma^{0}\phi^{+}-\frac{1}{\ell}\bar{\zeta}^{+}\gamma^{0}\psi^{+}-\frac{1}{\ell}\bar{\eta}\gamma^{0}\rho\,,\notag \\
\delta y&=&-\bar{\epsilon}^{-}\gamma^{0}\phi^{-}-\bar{\zeta}^{-}\gamma^{0}\psi^{-}-\bar{\epsilon}^{+}\gamma^{0}\phi^{+}-\bar{\zeta}^{+}\gamma^{0}\psi^{+}-\bar{\eta}\gamma^{0}\rho\,,\notag \\
\delta y_1&=&-\frac{1}{\ell}\bar{\epsilon}^{+}\gamma^{0}\psi^{+}\,,\notag \\
\delta y_2&=&+\frac{1}{\ell}\bar{\epsilon}^{-}\gamma^{0}\psi^{-}-\frac{1}{\ell}\bar{\epsilon}^{+}\gamma^{0}\rho-\frac{1}{\ell}\bar{\eta}\gamma^{0}\psi^{+}\,,\notag \\
\delta y_3&=&+\frac{1}{\ell}\bar{\epsilon}^{-}\gamma^{0}\phi^{-}+\frac{1}{\ell}\bar{\zeta}^{-}\gamma^{0}\psi^{-}-\frac{1}{\ell}\bar{\epsilon}^{+}\gamma^{0}\phi^{+}-\frac{1}{\ell}\bar{\zeta}^{+}\gamma^{0}\psi^{+}-\frac{1}{\ell}\bar{\eta}\gamma^{0}\rho\,,\notag \\
\delta u_1&=&-\bar{\epsilon}^{+}\gamma^{0}\psi^{+}\,,\notag \\
\delta u_2&=&+\bar{\epsilon}^{-}\gamma^{0}\psi^{-}-\bar{\epsilon}^{+}\gamma^{0}\rho-\bar{\rho}\gamma^{0}\psi^{+}\,,\notag \\
\delta u_3&=&+\bar{\epsilon}^{-}\gamma^{0}\phi^{-}+\bar{\zeta}^{-}\gamma^{0}\psi^{-}-\bar{\epsilon}^{+}\gamma^{0}\phi^{+}-\bar{\zeta}^{+}\gamma^{0}\psi^{+}-\bar{\eta}\gamma^{0}\rho\,,\label{gt}
\end{eqnarray}
along with
\begin{eqnarray}
\delta \psi^{+}&=&d\epsilon^{+}+\frac{1}{2}\omega\gamma_0\epsilon^{+}+\frac{1}{2\ell}\tau\gamma_0\epsilon^{+}-\frac{1}{2}y_1\gamma_0\epsilon^{+}\,\notag \\
\delta \psi^{-}&=&d\epsilon^{-}+\frac{1}{2}\omega\gamma_0\epsilon^{-}+\frac{1}{2\ell}\tau\gamma_0\epsilon^{-} +\frac{1}{2}\omega^{a}\gamma_a\epsilon^{+} +\frac{1}{2\ell}e^{a}\gamma_a\epsilon^{+}+\frac{1}{2}y_1\gamma_0\epsilon^{-}\,\notag \\
\delta \rho&=&d\eta+\frac{1}{2}\omega\gamma_0\eta+\frac{1}{2}s\gamma_0\epsilon^{+}+\frac{1}{2}\omega^{a}\gamma_a\epsilon^{-}+\frac{1}{2\ell}\tau\gamma_0\eta+\frac{1}{2\ell}m\gamma_0\epsilon^{+}+\frac{1}{2\ell}e^{a}\gamma_a\epsilon^{-}\notag \\
& &-\frac{1}{2}y_1\gamma_0\eta-\frac{1}{2}y_2\gamma_0\epsilon^{+}\,,\notag \\
\delta \phi^{+}&=&d\zeta^{+}+\frac{1}{2}\omega\gamma_0\epsilon^{+}+\frac{1}{2}s\gamma_0\eta+\frac{1}{2}\omega^{a}\gamma_a\zeta^{-}+\frac{1}{2}b^{a}\gamma_a\epsilon^{-}+\frac{1}{2\ell}\tau\gamma_0\zeta^{+}+\frac{1}{2\ell}m\gamma_0\eta\notag \\
& &+\frac{1}{2\ell}e^{a}\gamma_a\zeta^{-}+\frac{1}{2\ell}t^{a}\gamma_a\epsilon^{-}-\frac{1}{2}y_1\gamma_0\zeta^{+}-\frac{1}{2}y_2\gamma_0\eta-\frac{1}{2}y_3\gamma_0\epsilon^{+}\,,\notag \\
\delta \phi^{-} &=&d\zeta^{-}+\frac{1}{2}\omega\gamma_0\zeta^{-}+\frac{1}{2}s\gamma_0\epsilon^{-}+\frac{1}{2}\omega^{a}\gamma_a\eta+\frac{1}{2}b^{a}\gamma_a\epsilon^{+}+\frac{1}{2\ell}\tau\gamma_0\zeta^{-}+\frac{1}{2\ell}m\gamma_0\epsilon^{-}\notag \\
& &+\frac{1}{2\ell}e^{a}\gamma_a\eta+\frac{1}{2\ell}t^{a}\gamma_a\epsilon^{+}+\frac{1}{2}y_1\gamma_0\zeta^{-}+\frac{1}{2}y_2\gamma_0\epsilon^{-} +\frac{1}{2} z \gamma_0 \epsilon^+ + \frac{1}{2\ell} y \gamma_0 \epsilon^+ \,, \label{gt2}
\end{eqnarray}
where $\epsilon^{\pm}$, $\zeta^{\pm}$ and $\eta$ are the respective fermionic gauge parameters related to the fermionic generators $Q^{\pm}$, $W^{\pm}$ and $R$. One can notice that in the vanishing cosmological constant limit, $\ell\rightarrow\infty$, we recover the supersymmetry transformations laws under which the curvature two-forms of the extended Newtonian superalgebra transform in a covariant way \cite{Ozdemir:2019orp}.

\bibliographystyle{fullsort.bst}
 
\bibliography{Exotic_Newtonian_supergravity_vf}

\end{document}